\documentclass[lettersize,journal]{IEEEtran}
\usepackage{amsmath,amsfonts}
\usepackage{array}
\usepackage[caption=false,font=normalsize,labelfont=sf,textfont=sf]{subfig}
\usepackage{textcomp}
\usepackage{stfloats}
\usepackage{url}
\usepackage{verbatim}
\usepackage{graphicx}
\usepackage{cite}
\usepackage[table]{xcolor}
\definecolor{lightblue}{rgb}{0.85,0.9,1}
\usepackage{booktabs}  
\usepackage{colortbl}  
\usepackage{makecell}  
\usepackage{xcolor}
\usepackage{amsmath,amsfonts}  
\usepackage{algorithm}  
\usepackage{algorithmic}    
\usepackage{xspace}
\usepackage{multirow}
\usepackage{makecell} 
\begin{document}

\title{WebGeoInfer: A Structure-Free and Multi-Stage Framework for Geolocation Inference of Devices Exposing Information}

 \author{
Huipeng Yang, Li Yang, Lichuan Ma, Lu Zhou, Junbo Jia, Anyuan Sang, Xinyue Wang
\thanks{This work was supported in part by the National Key R\&D Program of
China (2023YFB3106900), country Natural Science Foundation of China
(62302362), the Fundamental Research Funds for the Central Universities
and the Innovation Fund of Xidian University(YJSJ25012).}        
\thanks{Huipeng Yang, Li Yang \textit{(Corresponding author)}, Lu Zhou, Junbo Jia, Anyuan Sang and  Xinyue Wang are with the School of Computer Science and Technology, Xidian University, Xi'an 710126, China, and also with the Shaanxi Key Laboratory of Network and System Security. Xi'an 710126, China.(e-mail:yanghp@stu.xidian.edu.cn, yangli@xidian.edu.cn, zhoulu@xidian.edu.cn, jbjia@stu.xidian.edu.cn, anyuan@stu.xidian.edu.cn, 21009290060@stu.xidian.edu.cn).}
\thanks{Lichuan Ma is with the School of Cyber Engineering. Xidian University, Xi'an 710126, China, and also with the Shaanxi Key Laboratory of Blockchain and Secure Computing, Xi'an 710126, China.(e-mail:lcma@xidian.edu.cn)}% <-this % stops a space
}

\markboth{Journal of \LaTeX\ Class Files,~Vol.~14, No.~8, August~2021}%
{Shell \MakeLowercase{\textit{et al.}}: A Sample Article Using IEEEtran.cls for IEEE Journals}


\maketitle

\begin{abstract}

Remote management devices facilitate critical infrastructure monitoring for administrators but simultaneously increase asset exposure. Sensitive geographical information overlooked in exposed device management pages poses substantial security risks. Therefore, identifying devices that reveal location information due to administrator negligence is crucial for cybersecurity regulation. Despite the rich information exposed by web interfaces of remote management devices, automatically discovering geographical locations remains challenging due to unstructured formats, varying styles, and incomplete geographical details.

This study introduces WebGeoInfer, a structure-free geolocation inference framework utilizing multi-stage information enhancement. WebGeoInfer clusters similar device web pages and analyzes inter-cluster differences to extract potential geographical information, bypassing structural limitations. Through search engine enhancement and Large Language Models mining, the framework extracts geographical coordinates from identified information. WebGeoInfer successfully inferred locations for 5,435 devices across 94 countries and 2,056 cities, achieving accuracy rates of 96.96\%, 88.05\%, and 79.70\% at country, city, and street levels, respectively.

\end{abstract}


\section{Introduction}

\IEEEPARstart{W}{ith} the rapid advancement of information technology, remote management devices have experienced exponential deployment growth across the public internet, providing system administrators with convenient and efficient service interfaces. However, due to misconfiguration or lack of security awareness, numerous device management pages are directly exposed on the internet. These web interfaces contain customizable sections where administrators can set personalized information for easier management, which often includes sensitive geographical data. This inadvertent information exposure significantly increases cybersecurity risks (such as targeted attacks against high-value assets), typically without the administrators' knowledge~\cite{sasaki2022exposed}. Regulatory authorities, as the final line of defense in cybersecurity governance, are responsible for identifying these devices exposing sensitive information and alerting their affiliated organizations. Consequently, geolocation inference of remote management devices has emerged as a critical research topic in cybersecurity~\cite{liu2022discover}, with the core objective of accurately mapping exposed devices in cyberspace to geographical coordinates in the physical world~\cite{zilberman2024survey}, providing essential support for device risk assessment and security policy formulation.

Existing geolocation inference methods for remote management devices can be classified into two major categories: active geolocation and passive geolocation. Active geolocation methods~\cite{ma2023graphnei,liu2025gdd,cho2024selection,zu2022ip} obtain network characteristics (such as latency and hop counts) by sending probe packets (e.g., Ping, Traceroute) to target devices, then infer target locations by combining this data with landmark information. However, these methods face significant limitations in our research context. First, they heavily depend on the quantity and accuracy of landmarks, yet the remote management device domain currently lacks large-scale landmark data. Second, active probing is susceptible to network fluctuations, which introduces instability in measurement results. Third, many remote management devices implement firewall policies that block probe packets, further limiting the application of active methods. In contrast, passive geolocation methods~\cite{lin2024probegeo,yan2022internet,cheng2021identify,dan2022ip} infer geographic locations by analyzing information already exposed by devices, without sending additional probe packets. The web interfaces of remote management devices expose rich information that provides excellent sources for passive geolocation, including detailed configurations set by administrators that often contain location information. This makes passive methods more suitable for our research scenario, as they can leverage existing public information while avoiding the limitations of active probing techniques.

While web page content analysis methods offer significant advantages for remote management device localization, current approaches still face several critical challenges. Existing Web page content analysis primarily extracts relevant information from device metadata~\cite{lin2024probegeo}, HTML content~\cite{cheng2021identify}, and embedded images~\cite{yan2022internet}. These approaches typically rely on rule-based extraction patterns or keyword matching techniques to identify target information. However, such methods encounter two major limitations in practical applications. 
\begin{itemize}
    \item \textbf{The information heterogeneity challenge:} different device manufacturers implement vastly different web interface designs, resulting in significant variations in how location information is represented—varying in format, position, and semantic expression across different devices. This heterogeneity makes it extremely difficult to design universal extraction rules that function effectively across various device types.
    \item \textbf{The information sparsity challenge:} location data in web interfaces is often fragmented and incomplete, with many devices only exposing partial information (such as abbreviated city names, acronyms of building names, etc.), requiring sophisticated inference to complete missing details.
\end{itemize}

These challenges collectively hinder the accuracy and reliability of existing passive geolocation methods based on Web page content analysis, necessitating more advanced approaches that can adaptively process diverse information formats while effectively mining authentic location information~\cite{dan2022ip}.

To address these challenges, we propose WebGeoInfer, a "web structure-free" geolocation inference framework that employs multi-stage information enhancement. This framework consists of three components: data collection and processing, web page differential text extraction, and geographic clue mining. 

The web page differential text extraction module addresses the web page heterogeneity problem by automatically identifying potential geographic location text. We observe that: different types of devices have varied page structures, but most web content of same-type devices is identical (for example, the same manufacturer information); While the differential text portions are typically administrator-customized content containing specific information such as geographic locations, we cluster pages based on DOM structure features to obtain sets of same-type device pages, then perform differential analysis within clusters to extract geographic clues. This method is fully automated, requires no manual adjustments for different page structures, and truly achieves "web structure-free" functionality.

For the information sparsity problem, we employ multi-stage information enhancement processing that leverages the semantic reasoning capabilities of online search engines. By utilizing the excellent text processing and reasoning analysis capabilities of Large Language Models (LLMs), we construct a comprehensive reasoning model that integrates the advantages of multiple LLMs. This model integrates contextual semantics to extract optimal geographic location information from search engine results. Additionally, we combine this approach with geographic information databases to improve the reliability of results.

To validate the effectiveness of our proposed method, we developed a system and conducted practical application tests. The results demonstrate that our solution successfully infers the geographic locations of 5,435 remote management devices distributed across 94 countries, 2,056 cities, and 336 autonomous system domains.

In summary, our work makes three main contributions:
\begin{enumerate}
\item \textbf{Web-Structure-Free text extraction method}: We propose an innovative page differential analysis technique that automatically extracts text containing geographic clues through page clustering and differential comparison. This method does not depend on specific web page structures or templates, adapting to the diverse web interfaces of various remote management devices.
\item \textbf{Multi-Stage-Enhanced inference framework}: We develop a deep semantic analysis inference framework that integrates search engines and LLMs. This framework can automatically extract and enrich geographic location information from incomplete and ambiguous text, achieving high-precision location inference through contextual semantic understanding.
% , significantly reducing human dependency and improving inference efficiency.
\item \textbf{Comprehensive experimental validation}: By building a prototype system and conducting real-world validation, we successfully obtained geographic location information for 5,435 remote management devices, achieving accuracy rates of 96.96\%, 88.05\%, and 79.70\% at the country, city, and street levels, respectively. These experimental results thoroughly demonstrate the reliability and practical value of our method in real-world application scenarios.
\end{enumerate}

The remainder of this paper is organized as follows: Section \ref{sec:Related Work} introduces related work on geographic location inference; Section \ref{sec:Motivation} presents our research motivation; Section \ref{sec:Framework} details our solution design; Section \ref{sec:Evaluation of framework} provides comprehensive experimental verification and result presentation; Section \ref{sec:Results Analysis} analyzes device geographic location inference results and provides insights into device network and geospatial distribution; Section \ref{Discussion} discusses the limitations of our method; and finally, Section \ref{Conclusion} concludes the paper.

\begin{figure}
  \includegraphics[width=\linewidth]{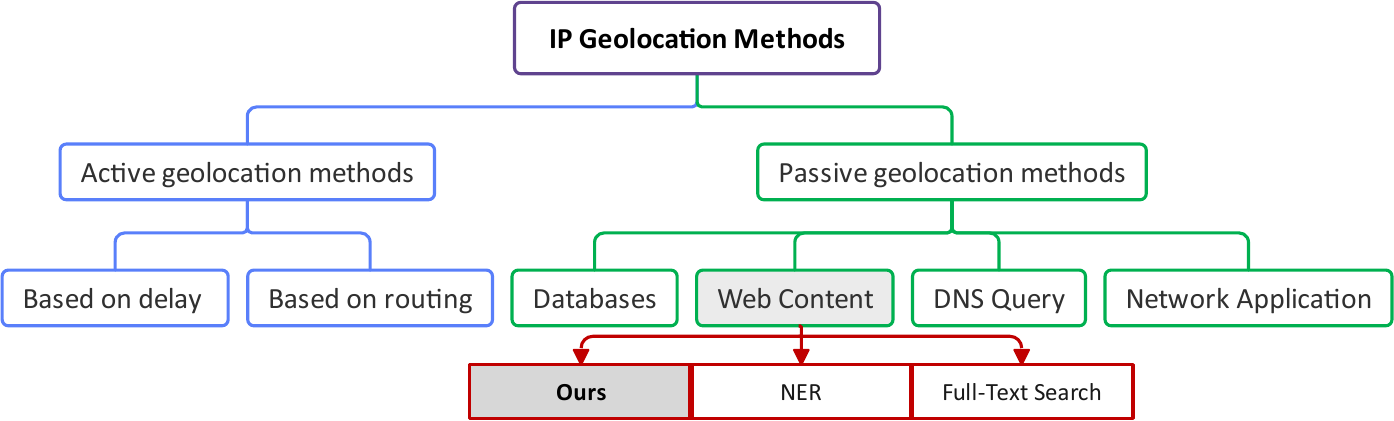}
  \caption{Classification of Cyberspace Device Geolocation Methods.}
  % \Description{Enjoying the baseball game from the third-base
  % seats. Ichiro Suzuki preparing to bat.}
  \label{fig:geolocation_classes}
\end{figure}

\section{Related Work}
\label{sec:Related Work}

Methods for inferring device geographical locations in cyberspace fall into two main categories: active and passive geolocation methods, as shown in Figure~\ref{fig:geolocation_classes}.
Given our research focus, we primarily analyze web content-based passive geolocation methods closely related to our work, while briefly introducing other approaches to provide comprehensive technical context.

\subsection{Active Geolocation Methods} 

Active geolocation methods involve using probing hosts with known locations to send probe packets to target devices. Traditional delay-based methods (such as SLG, CBG~\cite{gueye2004constraint}, and Spotter~\cite{laki2011spotter}) rely on the assumption of a positive correlation between network latency and geographic distance~\cite{wang2011towards}, while routing-based methods combine network topology to enhance location inference accuracy~\cite{xiang2019no,katz2006towards}. Recently, advancements in deep learning have significantly enhanced geographic inference capabilities, with approaches like GNN-Geo~\cite{ding2023gnn}, GraphGeo~\cite{wang2022connecting}, and NeighborGeo~\cite{wang2025neighborgeo} using graph neural networks to capture relationships between IP hosts. However, these methods either suffer from large location errors due to network noise or rely on substantial landmark data for training, presenting significant limitations.

\subsection{Passive Geolocation Methods}

This section will focus on analyzing web content-based methods, while briefly introducing other passive geolocation techniques.

\subsubsection{Web Content-based Methods}

Mansoori and Welch~\cite{mansoori2020they} conducted a comprehensive study on geolocation tracking techniques used by malicious websites, revealing various methods for obtaining location information from web elements. Zilberman et al.~\cite{zilberman2024survey} noted that extracting geographic information from device web interfaces represents an emerging and promising passive geolocation method. 

Methods for extracting geographic information from web content mainly include Named Entity Recognition (NER) and deep text retrieval. Berragan et al.~\cite{berragan2023transformer} built several NER models for extracting place names from unstructured text. Cruz et al.~\cite{cruz2013semantic} studied semantic extraction of geographic data from web tables. However, these methods face challenges with remote management device interfaces: NER methods require extensive training data and struggle with cross-domain generalization; deep text retrieval has high computational costs, limiting application in large-scale regulatory contexts.

\subsubsection{Other Passive Geolocation Methods}

Including database-based, DNS query-based, and network application-based methods.

Database-based Methods: Directly obtain information from Whois~\cite{endo2010whois}, RDAP~\cite{corneo2024whois}, CAIDA~\cite{gasser2016scanning}, or commercial geolocation services. Komosny et al.\cite{komosny2017location} shows these methods can achieve city-level positioning at best. While IPvSeeYou\cite{rye2023ipvseeyou} offers high precision, it's only applicable to EUI-64 IPv6 addresses.

DNS Query-based Methods: The Geo-Track algorithm~\cite{padmanabhan2001investigation} utilizes DNS hostnames to determine locations, the DRoP method~\cite{huffaker2014drop} improves accuracy by analyzing geographic hints in DNS data, and Dan et al.~\cite{dan2021ip} generate candidate location lists by integrating multiple data sources.

Network Application-based Methods: Mun and Lee~\cite{mun2017building} analyze online second-hand markets to build geographic databases, GeoBLR~\cite{du2019geoblr} employs Bayesian linear regression using user-shared data, and ONE-Geo~\cite{wang2019one} extracts owner names from web servers to obtain location information.

These methods are typically limited by data quality, coverage, or update frequency, presenting significant limitations when dealing with large-scale, dynamically changing network devices.

\textbf{In summary, existing geolocation methods have significant limitations:} active geolocation methods primarily rely on latency and routing information, are susceptible to network noise, or require extensive landmark data; passive geolocation methods are constrained by data source quality and availability. Although the remote management page exposes a large amount of device configuration information, existing web content-based methods mainly focus on extracting explicit geographic information from standardized web page elements, struggling to address the highly customized and non-standardized characteristics of remote management device interfaces. There remains a lack of systematic approaches to effectively extract geographic clues from non-standardized, highly customized remote management device web pages.

\section{Motivation}
\label{sec:Motivation}

Our study aims to develop an effective method for extracting geographic location information from web pages exposed by these devices. Specifically, achieving this goal faces two fundamental challenges:

    \textbf{\textit{1. How can we efficiently identify text containing geographic information from highly diverse and non-standardized device web pages?}}

    \textbf{\textit{2. How can we accurately infer geographic locations when geographic clues extracted from these pages are incomplete or ambiguous?}}

To address these challenges, we propose an innovative geographic inference framework for remote management devices.

\subsection{Information Extraction Based on Web Page Differences}

\begin{figure}[t]
  \includegraphics[width=\linewidth]{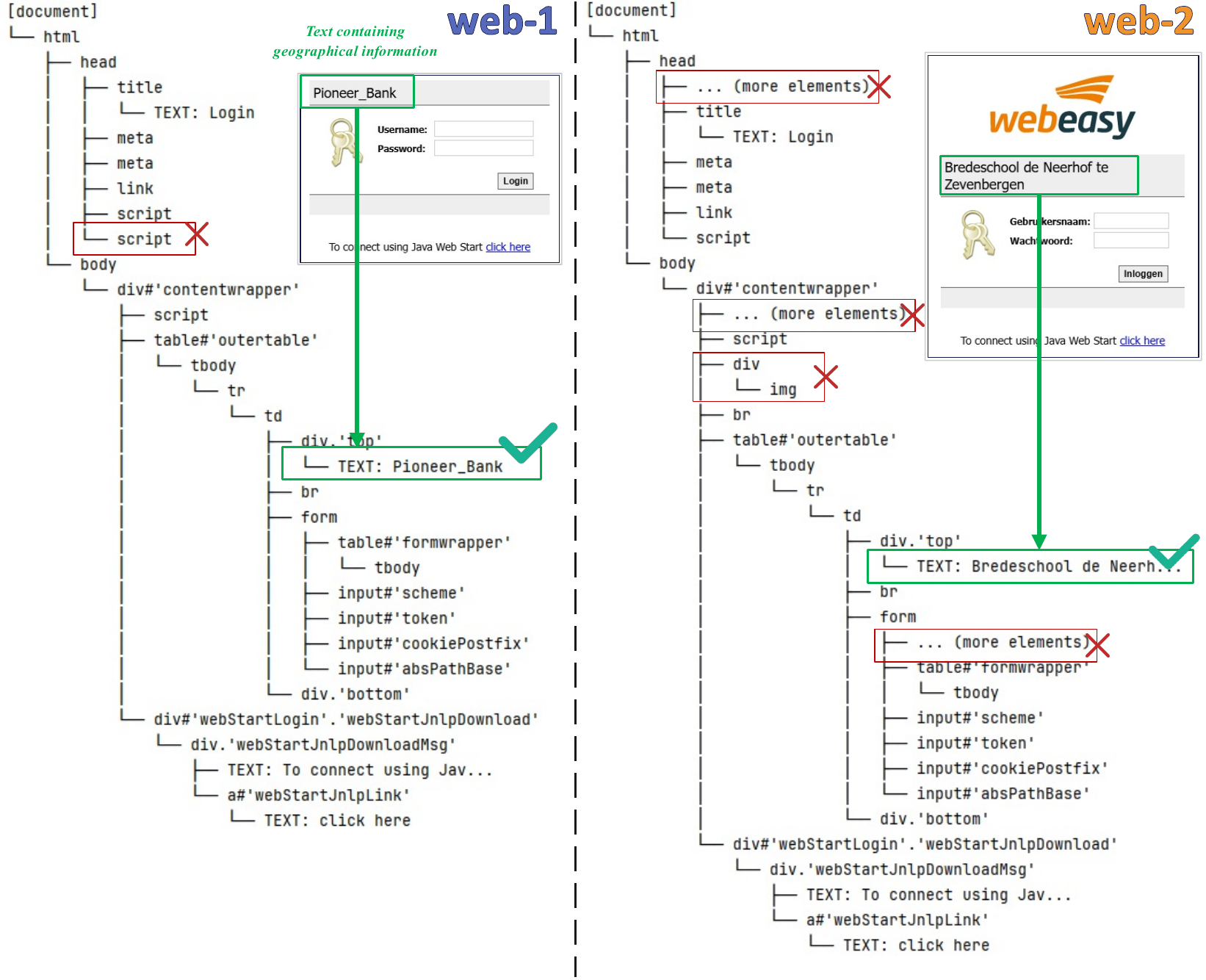}
  \caption{Comparing DOM tree differences across similar web pages.}
  % \Description{Enjoying the baseball game from the third-base
  % seats. Ichiro Suzuki preparing to bat.}
  \label{fig:web_different}
\end{figure}

Our analysis of numerous remote device management interfaces reveals that pages containing geographical location information are typically those with personalized configuration settings. While web page structures of similar devices are highly analogous, subtle differences exist due to customizations implemented by administrators for ease of management or identification. These subtle variations often contain high-value information, which corresponds to geographical location data in our study. Figure~\ref{fig:web_different} illustrates how these differences manifest primarily in two distinct categories:

\begin{itemize}
    \item \textbf{Structural Differences} (marked with red boxes): Variations at the HTML element level, including image elements that serve as corporate identifiers in specific deployments.
    \item \textbf{Content Differences} (marked with green boxes): Variations in textual information within identical structural components, encompassing device specifications, configuration parameters, and geographic deployment details—constituting the primary focus of our research.
\end{itemize}

This observation inspired us to develop an approach that functions independently of web page structure. Our method first achieves automated clustering of similar pages, and then extracts text content that may contain geographical information by comparing differences between similar web pages, thereby circumventing the limitations of traditional methods. 

\begin{figure*}[t]
  \includegraphics[width=\textwidth]{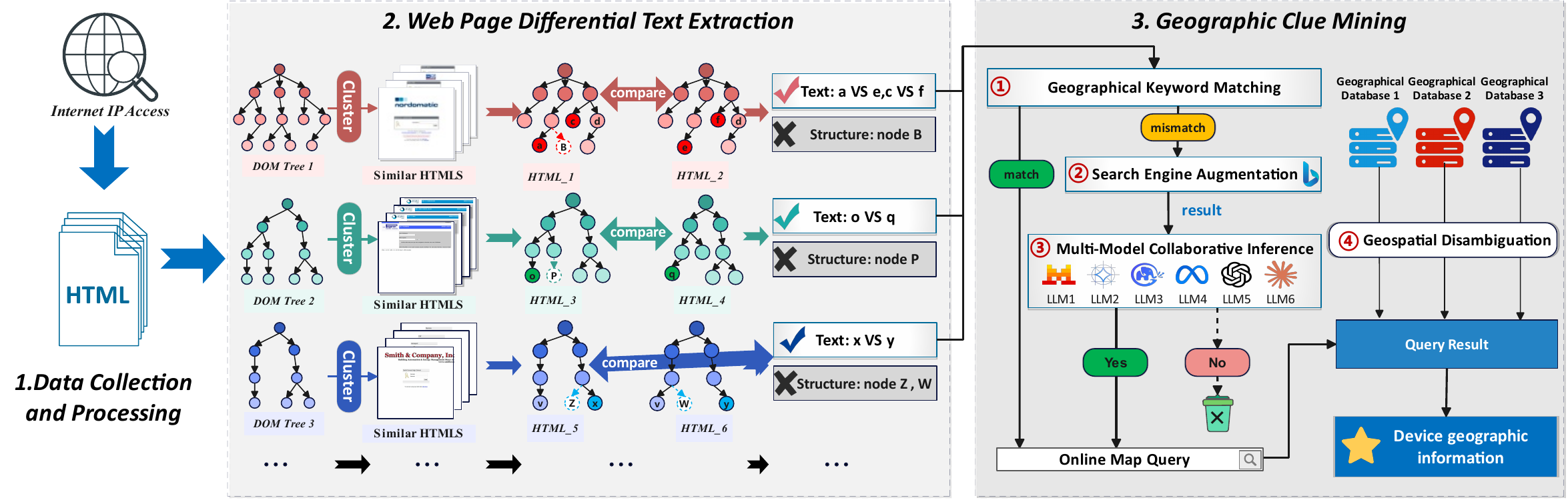}
  \caption{Overall framework. Including Data Collection and Processing, Web Page Differential Text Extraction and Geographic Clue Mining.}
  \label{fig:overall}
\end{figure*}

\subsection{Geographic Clue Mining and Enhancement}

Text extracted from web page differences frequently contains geographic indicators in abbreviated or implicit forms, rendering traditional NER techniques ineffective for direct application. To maximize the extraction of geographical clues, we introduce a multi-stage enhancement framework that leverages both the semantic enrichment capabilities of search engines and the collaborative reasoning abilities of LLMs. Our framework comprises:

\begin{itemize}
    \item \textbf{Keyword Matching}: Extracts results containing clear geographical identifiers for immediate use in online map queries. Content resistant to direct matching becomes the focus of subsequent enhanced analysis.
    \item \textbf{Search Engine Information Enhancement}: Employs search engines to transform brief textual fragments into semantically rich descriptions, thereby augmenting potential geographic contextual information.
    \item \textbf{Multi-Model Collaborative Inference}: Integrates multiple LLMs to extract geographic information from enhanced textual content. This orchestrated combination of LLMs with diverse scales and architectures enhances inference robustness while maintaining high accuracy~\cite{chang2024survey, harrod2024text, chen2025ipdb}.
    \item \textbf{Geographical Database-Assisted Enhancement}: Validates geographical location information extracted by large models through online map queries. When significant discrepancies arise - such as identical street names occurring in both the United States and the United Kingdom—the system implements auxiliary judgment using geographical databases to ensure result reliability.
\end{itemize}

Our approach combines differential text extraction with enhanced semantic parsing to address both challenges: efficiently identifying potential geographic information through structural and content differences, and accurately inferring precise locations through our multi-stage enhancement framework. This approach surpasses the structural dependencies and semantic limitations of traditional methods, providing an efficient and feasible solution for the large-scale geographical localization of remote devices.

\section{Framework}
\label{sec:Framework}

To systematically address these challenges, we propose a comprehensive framework called WebGeoInfer consisting of three core components: data collection and processing, web page differential text extraction, and geographic clue mining. Figure \ref{fig:overall} presents an overview of the framework.

\subsection{Data Collection and Processing}

Our study systematically collects web page information of remote management devices from the public internet through a comprehensive scanning methodology. Operating within a regulatory framework, we examine information exposed by these devices on the public internet. The data collection process encompasses the following essential steps:

\textbf{Network Probing}: Systematically scans IP address ranges to identify remote devices with network management interfaces. This process targets common management service ports while adhering strictly to network ethics guidelines, ensuring complete regulatory compliance.

\textbf{Page Capture} : Comprehensive information collection of target device pages, primarily including:
\begin{itemize}
\item HTML source code (Original HTML page information).
\item Page screenshots (Used for intuitive analysis of subsequent data annotation).
% \item DOM tree structure
\end{itemize}

\textbf{Data Preprocessing}: Utilizes the lxml and BeautifulSoup libraries for structured HTML processing, extracting and standardizing DOM trees to establish a robust foundation for subsequent clustering analysis. This preprocessing workflow ensures both data consistency and analytical integrity.

\subsection{Web Page Differential Text Extraction}

To extract geographic text information embedded within page differences, we present a method based on structural similarity clustering and differential text extraction. This approach comprises two key stages: device page clustering and differential text extraction (as illustrated in Algorithms \ref{alg:page_clustering} and \ref{alg:differential_text}).

\subsubsection{Device Page Cluster Stage}

We implement a Nilsimsa-based Locality-Sensitive Hashing (LSH) scheme to cluster pages according to their structural similarity. The system serializes each page's DOM tree, generating a 128-bit hash digest, with similarity quantified through Hamming distance. Through hierarchical aggregation, the method clusters pages with similarity exceeding threshold 
$\theta$, creating compact clusters that share a common structural skeleton—thus establishing a robust foundation for subsequent text alignment.

\begin{algorithm}[!htbp]
\caption{Nilsimsa-based Device Page Clustering}
\label{alg:page_clustering}
\begin{algorithmic}[1]
    \STATE \textbf{Input:} A set of pages $\mathcal{P}$, similarity threshold $\theta$
    \STATE \textbf{Output:} Clusters $\mathcal{C}$

    \STATE Initialize hash set $\mathcal{H} \leftarrow \emptyset$
    \FOR{each page $p_i$ in $\mathcal{P}$}
        \STATE Extract DOM tree $T_i$, serialize to $S_i$
        \STATE Compute Nilsimsa hash $h_i \leftarrow \mathrm{NilsimsaHash}(S_i)$
        \STATE $\mathcal{H} \leftarrow \mathcal{H} \cup \{h_i\}$
    \ENDFOR

    \STATE Build similarity matrix $M$
    \FOR{each pair $(h_i,h_j)$, $i \neq j$}
        \STATE Calculate distance $d \leftarrow \mathrm{HammingDistance}(h_i,h_j)$
        \STATE $M_{i,j} \leftarrow (128 - d)/128$
    \ENDFOR

    \STATE Initialize single-page clusters $\mathcal{C} \leftarrow \{\{p_1\},\ldots,\{p_n\}\}$

    \WHILE{there exist $C_a \neq C_b$ s.t. $\mathrm{AvgSim}(C_a,C_b)\ge\theta$}
        \STATE Select pair with maximum average similarity
        \STATE Merge clusters
        \STATE Update cluster collection $\mathcal{C}$
    \ENDWHILE

    \RETURN $\mathcal{C}$
\end{algorithmic}
\end{algorithm}

The clustering process encompasses four essential steps: DOM tree serialization, Nilsimsa hash generation, similarity matrix construction, and hierarchical clustering. The similarity threshold $\theta$ (default: 40) is calibrated through comprehensive experimental analysis.

\subsubsection{Differential Text Extraction Stage}
Within each cluster $C_k \in \mathcal{C}$, we execute structural alignment through subtree isomorphism search to precisely identify node groups across pages.

\begin{algorithm}  
\caption{Differential Text Extraction}  
\label{alg:differential_text}  
\begin{algorithmic}[1]  
    \STATE \textbf{Input:} Clusters $\mathcal{C} = \{C_1, \ldots, C_m\}$ from Algorithm~\ref{alg:page_clustering}  
    \STATE \textbf{Output:} Differential text collection $\Delta$ for each cluster  
    
    \STATE Initialize empty collection $\Delta \leftarrow \emptyset$
    
    \FOR{each cluster $C_k \in \mathcal{C}$}  
        \STATE Let $C_k$ contain pages $\{p_{k1}, p_{k2}, \ldots, p_{kn}\}$  
        \STATE Parse each page $p_{ki}$ into a DOM tree $T_{ki}$  
        \STATE Identify isomorphic subtrees across $T_{k1}, \dots, T_{kn}$ using subtree isomorphism  
        \STATE Obtain aligned node groups $G=\{g_1,\ldots,g_r\}$  
        \STATE Initialize differential text set $\Delta_k \leftarrow \emptyset$
        \FOR{each aligned node group $g_j \in G$}  
            % \STATE Extract text contents $\{text(g_j,p_{k1}), text(g_j,p_{k2}), \ldots, text(g_j,p_{kn})\}$  
            \STATE Extract text contents $\{text(g_j,p_{ki}) | i=1,2,\ldots,n\}$
            \STATE Normalize text (remove extra spaces, unify case, etc.)  
            \STATE $textSet \leftarrow \{text(g_j,p_{k1}), \ldots, text(g_j,p_{kn})\}$
            \IF{there exist different texts in $textSet$}
                \STATE $\Delta_k \leftarrow \Delta_k \cup textSet$
            \ENDIF  
        \ENDFOR  
        \STATE $\Delta \leftarrow \Delta \cup \Delta_k$
    \ENDFOR  
    \RETURN $\Delta$  
\end{algorithmic}  
\end{algorithm}

For each aligned node group, the system directly compares text content at identical structural positions. When the algorithm detects textual variations at equivalent structural positions, it identifies these differences as differential markers that frequently reveal geographical location clues. This direct comparison methodology circumvents complex similarity calculations, thereby enhancing efficiency and facilitating the capture of subtle yet critical textual variations.

\subsection{Geographic Clue Mining}

This section presents our core framework for extracting geographic location intelligence from differential text segments. To systematically address the challenges in differential texts that contain non-geographic elements (device attributes, status indicators), or geographic information presented in abbreviated or implicit forms, we implemented a progressive four-stage mining process, including: Geographic Keyword Matching, Search Engine Information Augmentation, Multi-Model Collaborative Inference, and Geospatial Disambiguation with Database Assistance. Each stage builds upon the previous, creating an integrated pipeline that progressively refines geographical inference.

\subsubsection{First Stage-Deterministic Geographic Keyword Matching}

From the extracted text differences on the web page, clear geographic information can be identified, such as "Gilbert High School" and "Georgia Intercountry Convention Center," where "School" and "Center" explicitly indicate geographic location clues. To reduce subsequent model mining costs, in the first stage, we adopt a keyword matching method to quickly identify texts containing geographic clues. We extracted a total of 65 keywords, covering categories such as types of roads, types of buildings and places, commercial institutions, administrative and community places, and administrative divisions. It is noteworthy that these keywords were modified and defined based on existing standards (such as ISO 19112) to ensure they meet our research needs and enhance the accuracy and consistency of the identification. Detailed keyword information can be found in Table \ref{tab:keywords}. If these keywords are matched in the difference text, the text is considered to exhibit strong geographic location characteristics and is directly identified as geographic location clues.

\subsubsection{Second Stage-Search Engine Information Augmentation}

This stage addresses the challenge of incomplete geographic signifiers—particularly abbreviated forms, acronyms, and contextually dependent references that resist direct identification. When deterministic matching fails to yield results, the system implements a semantic expansion methodology that leverages commercial search engine capabilities. This technique systematically transforms low-entropy textual fragments into contextually enriched representations by querying and analyzing publicly accessible knowledge bases. The algorithm submits the differential text segments as search queries, then applies natural language processing techniques to extract and consolidate the geographic context from search results. This semantic enrichment process significantly enhances information density and geographical context, creating a robust foundation for the subsequent application of inference models. The methodology effectively bridges the gap between abbreviated references and their complete geographical significance.

\begin{table}[t]  
\centering  
\caption{Keyword Categories for Deterministic Geographic Matching}  
\setlength{\arrayrulewidth}{0.5pt}  
\renewcommand{\arraystretch}{1.1}  
\begin{tabular}{>{\arraybackslash\bfseries}m{0.3\linewidth}|>{\arraybackslash}m{0.6\linewidth}}  
\midrule
\textbf{Category} & \textbf{Specific Types} \\
\midrule 
Common Road Types & street, avenue, road, boulevard, highway, lane, way,   
drive, path, expressway, parkway, alley \\
\hline  
Venue and Business Types & market, shop, store, boutique, supermarket, pharmacy,   
bank, library, museum, restaurant, hotel, inn, resort, mall, plaza, center, square, park, garden, stadium, theater, cinema, arena, club \\
\hline  
Administrative and Community Places & city\_hall, courthouse, police\_station,  
fire\_station, community\_center \\
\hline  
Transportation Venues & airport, station, terminal, bus\_stop, subway, metro \\
\hline  
Special Types & bridge, castle, monument, landmark, district, neighborhood \\
\hline  
Administrative Divisions & city, town, village, county, state,   
province, region, municipality \\
\hline  
Country and Continental Levels & country, nation, continent, territory \\
\midrule
\end{tabular}  
\label{tab:keywords}
\end{table}

\subsubsection{Third Stage-Multi-Model Collaborative Inference}

In this stage, we utilize LLMs to extract geographic intelligence from search engine results through a methodical analytical process.  LLMs demonstrate exceptional capability in both explicit pattern recognition and implicit semantic relationship inference, enabling sophisticated geographical reasoning by integrating textual data with contextual knowledge. Our framework utilizes a heterogeneous model ensemble comprising open-source models and proprietary models, selected through comprehensive evaluation of geographical inference performance, domain-specific capabilities, and computational efficiency parameters.

We design systematic prompting protocols that implement Chain-of-Thought (CoT) reasoning techniques to enhance geographical inference precision: 

\begin{itemize}
    \item We implement entity recognition prompts: "Please analyze whether the following text contains geographic location information and extract specific geographic location names." guiding models toward semantic understanding. 
    \item We deploy contextual verification prompts: "Please determine whether the extracted geographic location information is accurate based on the context and add relevant details." to validate accuracy. 
    \item For ambiguity resolution, we apply inference-chain prompts: "If there is ambiguous geographic location information in the text, please infer its possible meaning and provide reasoning results." enhancing analysis robustness.
\end{itemize}

For resolving model disagreements, we implement a weighted ensemble decision system that computes confidence scores based on each model's demonstrated accuracy in geographical entity recognition. We formalize this approach as:

\begin{equation}
G^* = \arg\max_{G \in \mathcal{G}} \sum_{i=1}^{n} w_i \cdot c_i(G)
\end{equation}

where $G^*$ represents the final geographical inference result, $w_i$ is the confidence weight assigned to model $i$ based on its historical accuracy, and $c_i(G)$ is the confidence score that model $i$ assigns to the geographical entity $G$. This ensemble approach optimizes inference reliability by leveraging complementary strengths across diverse model architectures. The system then transforms extracted geographical entities into standardized coordinate representations through integration with geospatial mapping services.

\subsubsection{Fourth stage-Geospatial Disambiguation with Database Assistance}

Considering that online map queries of geographic information typically yield multiple candidate locations, we integrate IP geolocation databases to provide auxiliary disambiguation support.

For geographic location information acquired through mining, online map queries frequently return multiple possible coordinate positions, creating geographic reference ambiguity. For example, street names like "State Street" exist extensively across numerous American cities. In these cases, we apply IP geolocation databases to disambiguate the extracted geographic information, requiring constraint verification only at city or provincial levels. Previous research~\cite{du2020ripe,gharaibeh2017look,electronics13010015} demonstrates that integrating multiple IP geolocation databases provides reliable city-level positioning results.

The system analyzes the mapping relationship between device IP addresses and geographic space to construct regional constraint conditions, thereby identifying the most reasonable geographic correspondences among multiple candidate locations. This disambiguation mechanism significantly enhances the credibility of geographic inference results, particularly for cases with identical names but different geographic locations.

\section{Evaluation of framework}
\label{sec:Evaluation of framework}

\subsection{Datasets}

\subsubsection{\textbf{Original Corpus}}

As regulators, we require comprehensive internet scanning data for our analysis. The Censys dataset~\cite{censys15} provides extensive results from methodical low-speed scans across the global network infrastructure. We target data marked with "SCADA" identifiers, representing industrial control and remote management devices, creating an initial corpus of 205,028 distinct IP and port combinations. Our review reveals many endpoints implement authentication requirements, resulting in HTML files with insufficient content for analysis. This security feature shows these protected devices fall outside our regulatory focus. After content filtration, we establish a refined corpus of 89,477 HTML pages with sufficient information for analysis.

\subsubsection{\textbf{Clustering Threshold Tuning Dataset}}
\label{Clustering Threshold Tuning Dataset}
To determine an optimal Hamming distance threshold for Nilsimsa-based clustering of remote management device interfaces, we manually identify 54 distinct page classes from the 89,477-page corpus that potentially contain geographic information markers. These classes are selected specifically because they demonstrate consistent patterns of embedded geographic location data within their interface structures. Using these representative page templates, we identify and extract 10,709 structurally similar pages to construct a specialized device type dataset. This curated collection enables precise calibration of clustering similarity thresholds essential for accurate device categorization.

\subsubsection{\textbf{Gold Standard}}
\label{gold standard}

For system training and evaluation purposes, we manually construct a comprehensive \textit{gold standard} dataset. From the 10,709-page threshold tuning dataset, we methodically select pages containing identifiable geographic location indicators. This reference dataset incorporates multiple essential components: original HTML pages sourced directly from remote management devices, extracted differential text segments that distinguish individual devices, corresponding search engine query results derived from these differential text elements, and verified geographic location information extracted from these query results.

We implement a rigorous annotation protocol, assigning each candidate page to multiple volunteers and ensuring independent evaluation by at least two annotators per page. Only pages with complete inter-annotator agreement are retained in the final dataset. This labor-intensive curation process ultimately yields a gold standard collection comprising 1,408 fully validated entries that serve as our definitive evaluation benchmark.

\subsection{Experiment Setup}  

\subsubsection{\textbf{Web Page Differential Text Extraction}}
We utilize BeautifulSoup to parse HTML content, extracting text and DOM tree features which are concatenated to form a composite string for Nilsimsa hash generation. Similarity between pages is calculated using Hamming distance with a threshold of 40 (analysis in Section \ref{sec:parameter_analysis}). Our clustering algorithm iterates through hash values, assigning each page to an existing cluster or creating a new one. The first page added determines each cluster's representative hash. For differential text extraction, we select one reference file per cluster and compare remaining files against it.

\subsubsection{\textbf{Geographical Clue Mining}}
In the geographic keyword matching stage, we employ a list of 65 keywords (As shown in Table \ref{tab:keywords}) with case-insensitive matching to maximize identification accuracy.

In the search engine information augmentation stage, we implement programmatic control of Google Chrome using Selenium with ChromeDriver for automatic browser operation. Bing serves as our primary search engine due to its comprehensive result summaries, related images, and regional prioritization features that enhance geographic information extraction. Bing's interface also demonstrates better resistance to anti-automation controls, making it suitable for batch query operations.

In the multi-model collaborative inference stage, We evaluate both open-source models (Mistral, Gemma2, GLM4, LLaMA3.1) and closed-source models (ChatGPT-4o, Claude 3.7) for our framework.  We select Mistral-7b, GLM4-9b, Gemma2-9b, and LLaMA3.1-8b specifically for offline deployment compatibility with the Ollama tool, while accessing ChatGPT-4o and Claude 3.7 through online APIs. All models are evaluated using the \textit{gold standard} dataset from Section \ref{gold standard}.

In the geospatial disambiguation with database assistance stage, we integrate location data from IP2Location, IPinfo, MaxMind, and IPIP platforms. When discrepancies arise among database results, we apply majority voting to determine the definitive geographic location.

\subsection{Evaluation Metrics}  

For HTML page clustering effectiveness, we consider the following evaluation metrics:  

\begin{equation}  
    \text{Purity} = \frac{1}{N} \sum_{i=1}^{k} \max_j |C_i \cap T_j|  
\end{equation}  
where $N$ represents the total number of samples, $k$ denotes the number of clusters, $C_i$ represents the $i$-th cluster, and $T_j$ represents the $j$-th true class.  

{\small
\begin{equation}  
    \text{ARI} = \frac{\sum_{ij} \binom{n_{ij}}{2} - \left[ \sum_{i} \binom{a_i}{2} \sum_{j} \binom{b_j}{2} \right] / \binom{N}{2}}{\frac{1}{2} \left[ \sum_{i} \binom{a_i}{2} + \sum_{j} \binom{b_j}{2} \right] - \left[ \sum_{i} \binom{a_i}{2} \sum_{j} \binom{b_j}{2} \right] / \binom{N}{2}}  
\end{equation}  
}
where $n_{ij}$ represents the number of samples that belong to both cluster $i$ and class $j$, $a_i$ denotes the number of samples in cluster $i$, $b_j$ denotes the number of samples in class $j$, and $N$ is the total number of samples.  

\begin{equation}  
    \text{NMI}(C,T) = \frac{I(C,T)}{\sqrt{H(C)H(T)}}  
\end{equation}  
where $I(C,T)$ represents the mutual information between the clustering result $C$ and true classes $T$, and $H(C)$ and $H(T)$ are the entropies of $C$ and $T$, respectively.  

\begin{equation}  
    \text{Homogeneity} = 1 - \frac{H(T|C)}{H(T)}  
\end{equation}  

\begin{equation}  
    \text{Completeness} = 1 - \frac{H(C|T)}{H(C)}  
\end{equation}  
where $H(T|C)$ represents the conditional entropy of the true classes $T$ given the clustering result $C$, and $H(C|T)$ represents the conditional entropy of the clustering result $C$ given the true classes $T$.  

\begin{equation}  
    \text{V-measure} = \frac{2 \times \text{Homogeneity} \times \text{Completeness}}{\text{Homogeneity} + \text{Completeness}}  
\end{equation}  

For the geographic location information extraction capability of LLMs, we use the following metrics:  

\begin{equation}  
    \text{Coverage} = \frac{|P_{geo}|}{|P_{total}|} \times 100\%  
\end{equation}  
where $P_{geo}$ represents the set of pages with extracted geographic location information, and $P_{total}$ represents the set of all pages containing geographic location information.  

\begin{equation}  
    \text{Accuracy} = \frac{|G_{correct}|}{|G_{extracted}|} \times 100\%  
\end{equation}  
where $G_{correct}$ represents the set of correctly extracted geographic locations, and $G_{extracted}$ represents the set of all extracted geographic locations.

\subsection{Experiment Result}

\begin{table*}[t]
\centering
\caption{Performance Comparison of Different Geolocation Methods}
\label{tab:geolocation_comparison}
\renewcommand{\arraystretch}{1}
\resizebox{0.7\textwidth}{!}{
\begin{tabular}{c|cccc}
\toprule
\textbf{Evaluation Metric} &
\textbf{Whois Query~\cite{endo2010whois}} &
\textbf{rDNS Analysis~\cite{dan2021ip}} &
\textbf{ProbeGeo(NER)~\cite{lin2024probegeo}} &
\textbf{Our Method} \\ \midrule
Country-level Coverage & 79.10\% & 6.80\% & 14.60\% & \textbf{99.71\%} \\
City-level Coverage & 6.38\% & 6.78\% & 9.92\% & \textbf{99.71\%} \\
Street-level Coverage & -- & -- & 9.78\% & \textbf{98.40\%} \\ \midrule
Country-level Accuracy & 51.95\% & 1.59\% & 3.05\% & \textbf{96.96\%} \\ 
City-level Accuracy & 3.69\% & 0.28\% & 0.95\% & \textbf{88.05\%} \\ 
Street-level Accuracy & -- & -- & 0.19\% & \textbf{79.70\%} \\ \bottomrule
\end{tabular}
}
\end{table*}

\subsubsection{\textbf{Overall Performance}}  

Given the absence of existing research that perfectly aligns with our research objective (extracting geographic information from web pages), we construct a comparative experiment to systematically evaluate our proposed method against several mainstream passive geolocation techniques. The evaluation methods include traditional Whois query techniques~\cite{endo2010whois}, rDNS analysis methods~\cite{dan2021ip}, and the ProbeGeo~\cite{lin2024probegeo} framework representing the current state of the art. We particularly focus on the Device Landmark Generator component in the ProbeGeo framework, as its technical approach (using NER to extract geographic information from web page elements) shares similarities with our research objective. The evaluation dimensions include coverage rate and accuracy at different granularities (country, city, and street levels).

Table~\ref{tab:geolocation_comparison} presents the performance comparison results of various methods. Overall, there are significant differences among the evaluated methods. Specifically, the multi-stage enhancement method proposed in this paper demonstrates clear advantages across all evaluation metrics, with both coverage and accuracy significantly surpassing those of other methods, showcasing its effectiveness in handling geographic information extraction tasks.

The Whois query method shows relatively good coverage (79.10\%), but its accuracy is low, with a country-level accuracy of only 51.95\% and an even lower city-level accuracy of just 3.69\%. This result aligns with the fundamental characteristics of the Whois database, which records the registration addresses of IP address block administrative entities (typically ISPs) rather than the actual deployment locations of devices. Although Whois data holds some value for network management and legal tracing, it has inherent limitations in precise geolocation.

The rDNS analysis method faces challenges related to data availability. Our tests indicate that only 37.4\% of IP addresses can retrieve reverse DNS records, and only 6.80\% of those records contain usable geographic information. This is primarily because reverse DNS records are non-mandatory and depend on voluntary implementation by network administrators. Even when configured, the diversity of naming conventions complicates standardized geographic information extraction.

In terms of directly extracting geographic location information from HTML pages, the ProbeGeo framework has a coverage rate of only 14.60\%. This highlights the limitations of rule-based NER methods when dealing with highly unstructured data. Geographic information in HTML often appears in various forms (such as abbreviations, or incomplete expressions) that exceed the recognition capabilities of traditional NER models. Relying solely on entity recognition without deep semantic understanding presents significant bottlenecks in precise inference.

In contrast, our method demonstrates significant advantages across all evaluation metrics. Specifically, it achieves coverage rates of 99.71\% at city levels, and 98.40\% at the street level; in terms of accuracy, it reaches 96.96\%, 88.05\%, and 79.70\% at the country, city, and street levels, respectively. These outstanding results stem from the ability of search engines to infer and expand on text information, allowing ambiguous or incomplete geographic clues to be associated with broader geographic knowledge bases, thus providing relevant information augmentation. Additionally, the strong contextual and semantic understanding capabilities of LLMs enable effective and precise parsing of latent geographic information.

The comparative results indicate that methods based on Whois queries and rDNS analysis struggle to meet the demands for accurate geographic location information extraction in regulatory scenarios, further confirming the critical impact of passive geolocation data source selection on positioning effectiveness. Additionally, the comparison with rule-based NER methods validates that LLMs significantly outperform simple rule-matching approaches in semantic analysis through contextual understanding.

Furthermore, the trend of declining overall accuracy across various methods as geographic granularity shifts from country-level to street-level reflects the increasing complexity and difficulty inherent in geolocation tasks.

\begin{figure*}[t]  
    \centering  
    \subfloat[]{  
        \includegraphics[width=0.47\linewidth]{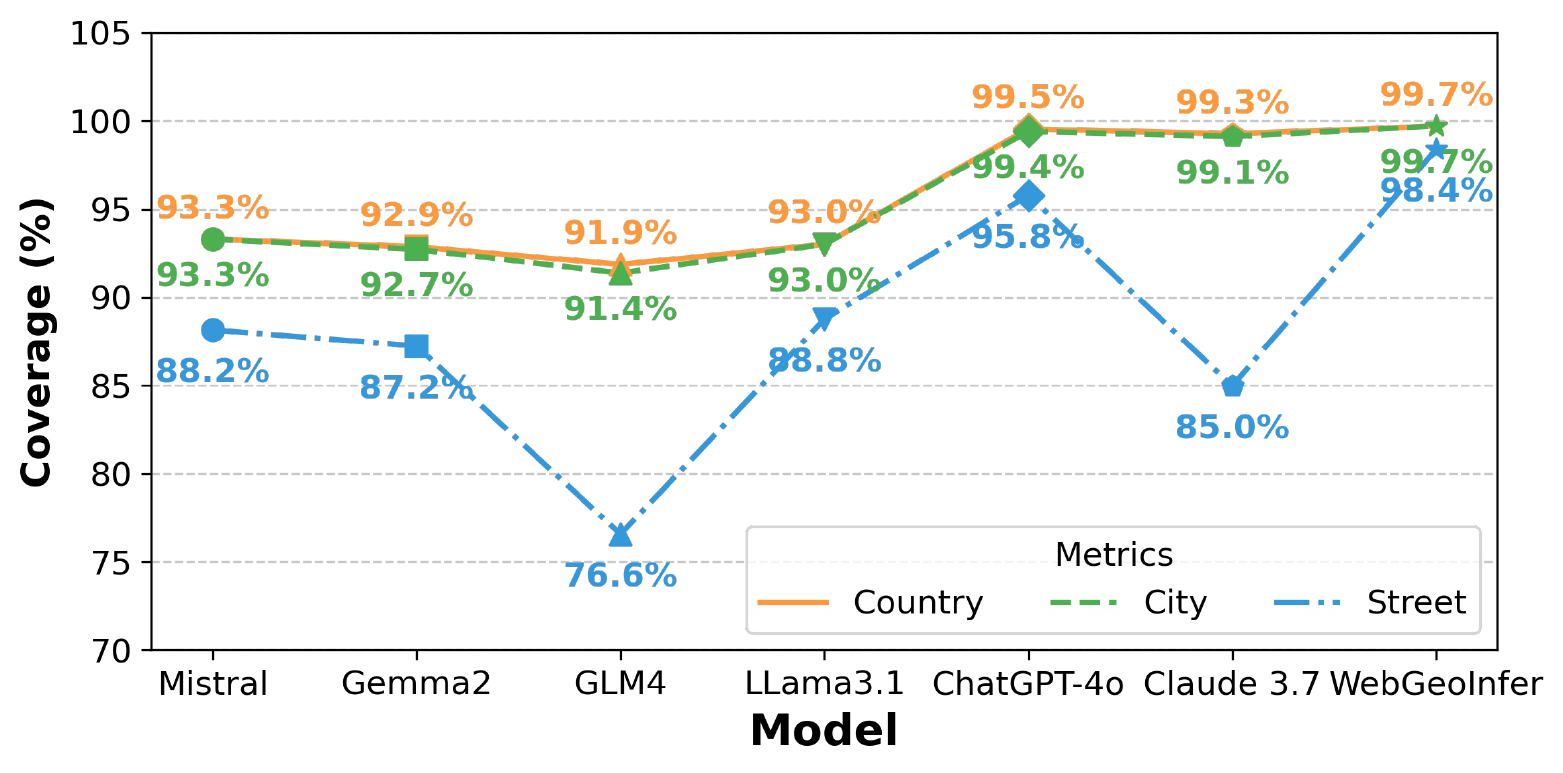}%  
        \label{fig:model_coverage_compare}  
    }  
    \hfil  
    \subfloat[]{  
        \includegraphics[width=0.47\linewidth]{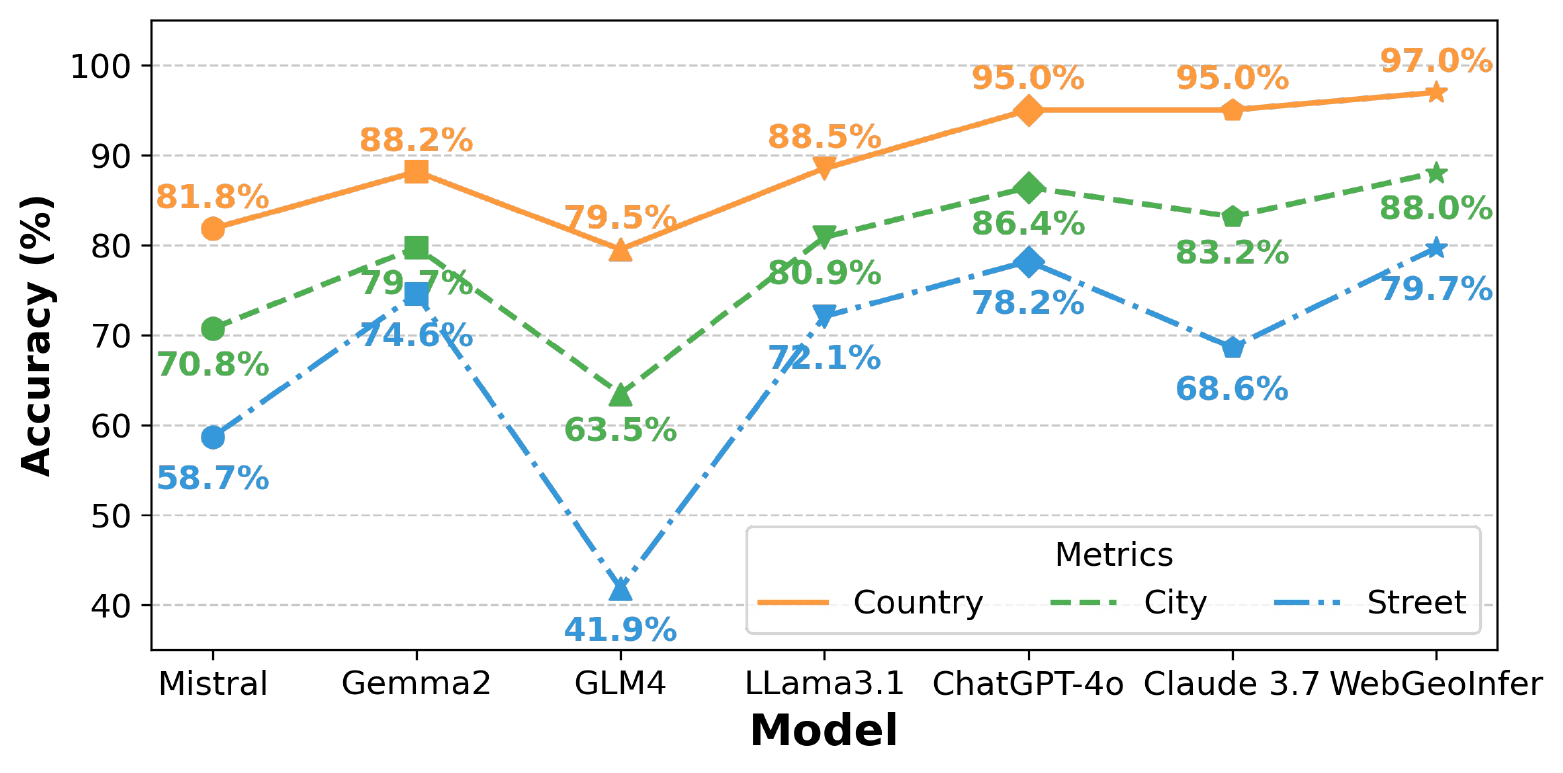}%  
        \label{fig:model_accuracy_compare}  
    }  
    \caption{Comparison of Large Language Model Performance. (a) shows the coverage of geographical location inference by different LLMs at the country, city, and street levels, while (b) shows their accuracy at these levels.}  
    \label{fig:model_compare}  
\end{figure*}

\subsubsection{\textbf{Comparison of Large Language Model Performance}}

To evaluate the effectiveness of different LLMs within our remote device geolocation inference framework, we first conduct experiments on a gold standard dataset, selecting 100 samples to test the models' accuracy. Based on their performance in terms of accuracy, we assign weights from highest to lowest: the online LLMs ChatGPT-4o and Claude 3.7 perform the best, receiving a weight of 2, while LLaMA 3.1 has the best overall performance among offline models, receiving a weight of 1.5. Other offline models are assigned a weight of 1.

Subsequently, we conduct experiments on the entire sample to evaluate the coverage and accuracy performance of these models at the country, city, and street levels.
As shown in Figure~\ref{fig:model_compare}, different LLMs exhibit significant performance variance. In coverage, ChatGPT-4o and Claude 3.7 both identify over 99\% of country- and city-level locations; notably, ChatGPT-4o achieves 95.79\% coverage at the street level. 
% In contrast, GLM4 lags behind, with only 76.57\% street-level coverage.

Accuracy differences are particularly evident across different levels. At the country level, both ChatGPT-4o and Claude 3.7 achieve an accuracy of 95.03\%, significantly surpassing other models. ChatGPT-4o achieves highest accuracy in geolocation extraction, with 86.42\% at city-level and 78.20\% at the more challenging street-level, while GLM4 reaches only 41.87\% at street-level, revealing its limitations in fine-grained extraction.

Our ensemble method outperforms the best single model performance across all metrics. Notably, our approach yields the greatest improvement in the most challenging street-level extraction, an area where individual models struggle the most. The consistent improvements across all levels of granularity validate the effectiveness of leveraging complementary strengths from multiple models.

\begin{figure}[t]
    \centering
    \includegraphics[width=0.95\linewidth]{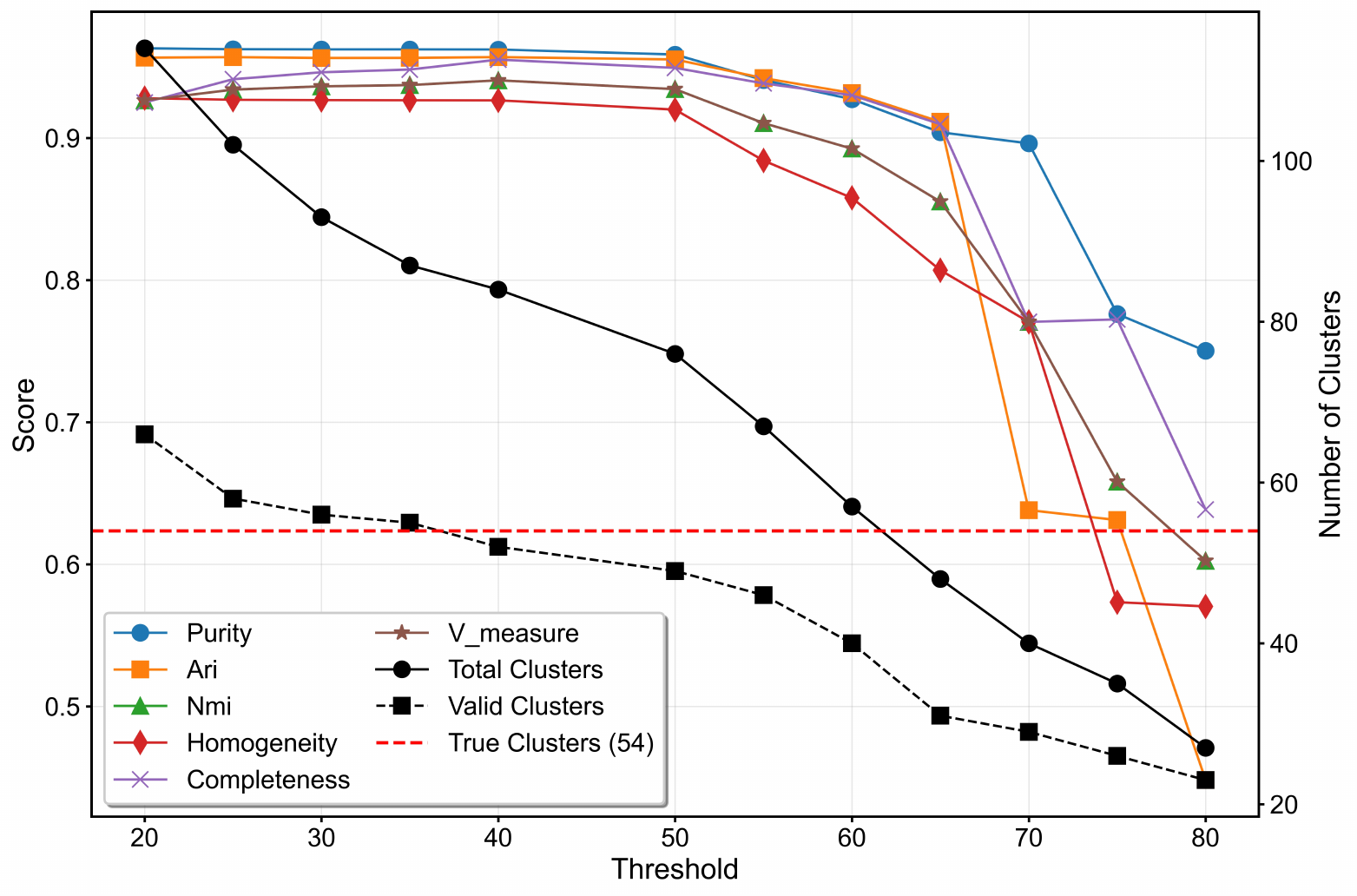}
    \caption{Clustering evaluation metrics under different thresholds, including Purity, ARI, NMI, Homogeneity, Completeness, and V-Measure. Total Clusters represents the original number of clusters, Valid Clusters represents the number of effective clusters after removing noise clusters, and True Clusters represents the ground truth number of clusters.}
    \label{fig:clustering_performance}
\end{figure}

\subsubsection{\textbf{Hyperparameter Experiment (Analysis of Different Threshold Settings in Differential Text Extraction)}}
\label{sec:parameter_analysis}

We evaluate clustering performance across threshold values from 20 to 80  using the dataset constructed in Section~\ref{Clustering Threshold Tuning Dataset}. A threshold of 40 is ultimately chosen because it offers the best balance between homogeneity and completeness. At this threshold, 83 clusters are formed: 52 clusters (size$\geq$ 2) closely align with 54 reference categories, while the remaining 32 singleton clusters (accounting for approximately 3.1\% of the total data) are treated as noise. The V‑measure reaches its highest value of 0.940581.

\begin{figure}[t]  
    \centering  
    \subfloat[]{  
        \includegraphics[width=0.46\linewidth]{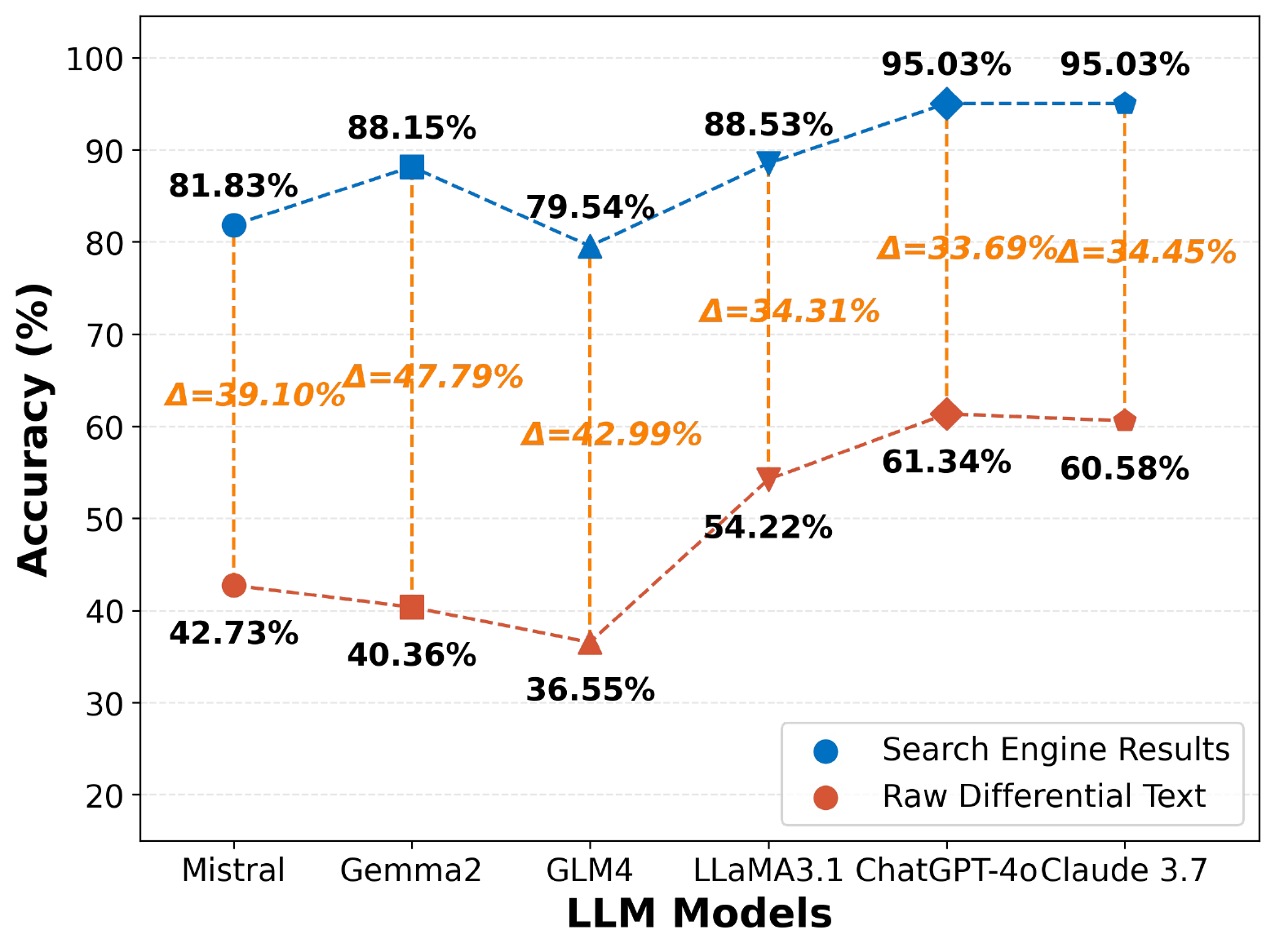}%  
        \label{fig:ablation_country}  
    }  
    \hfil  
    \subfloat[]{  
        \includegraphics[width=0.46\linewidth]{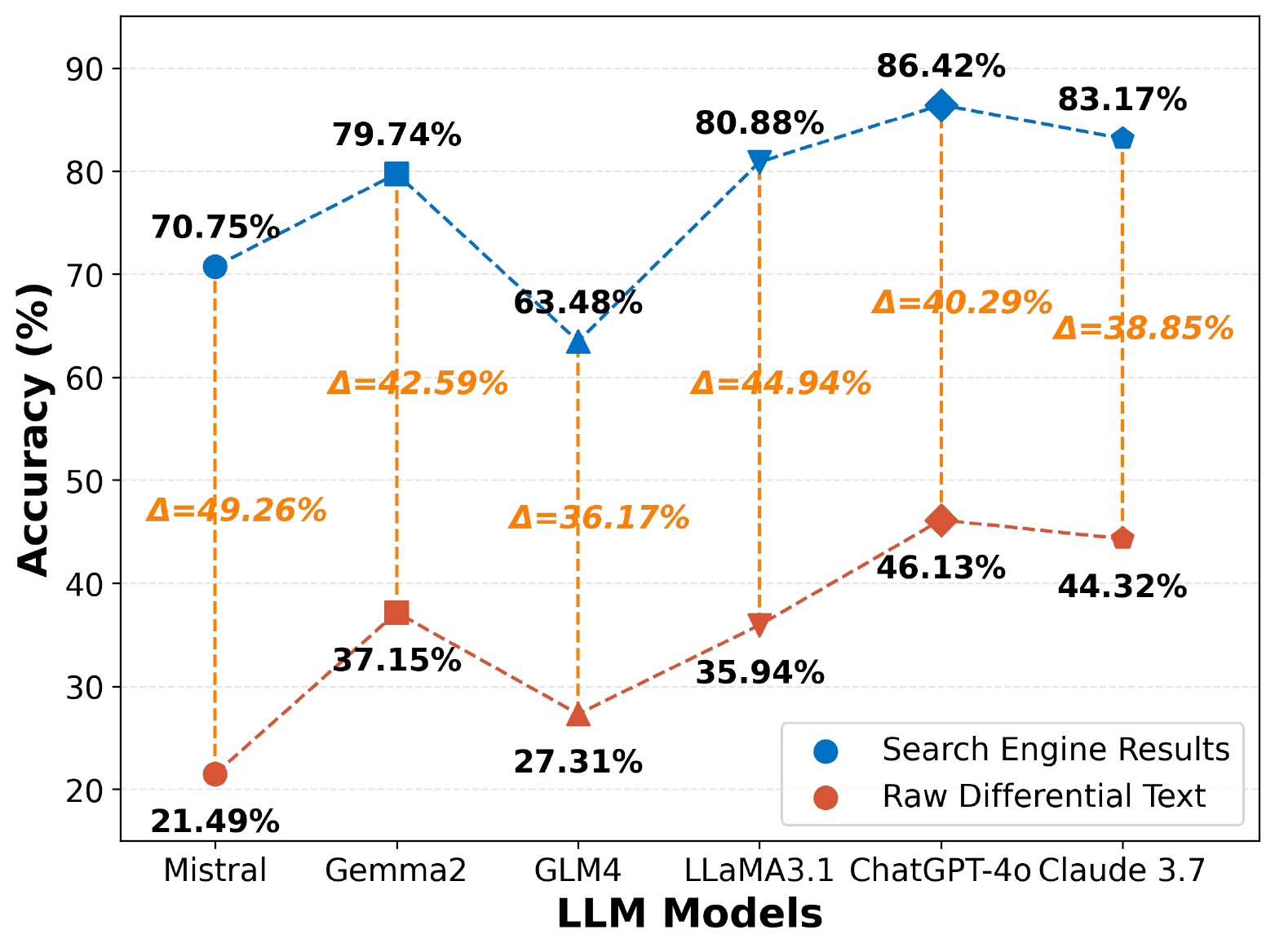}%  
        \label{fig:ablation_city}  
    }  
    \caption{Accuracy Comparison of LLMs on Raw differential text vs. Search Engine Query Results. (a) Gradient Stacked Area Chart of Country-Level Accuracy, (b) Gradient Stacked Area Chart of City-Level Accuracy.}  
    \label{fig:ablation_study}  
\end{figure}

\subsubsection{\textbf{Ablation Experiment (Search Engine Information Augmentation Analysis)}}  

To validate the effectiveness of information enhancement based on search engines within our framework, we design and implement ablation experiments. These experiments aim to compare performance under the following two scenarios:

\begin{itemize}
    \item  Directly identifying geographic location clues from the original differential text.
    \item  Geographic location clue identification after augmenting the original differential text with search engine query results.
\end{itemize}

We evaluated performance differences between methods using country and city-level accuracy metrics (Fig.\ref{fig:ablation_country} and Fig.\ref{fig:ablation_city}). Our ablation experiments show that search engine enhancement significantly improved geographic inference performance, increasing country-level accuracy by 34.31\%-47.79\% and city-level accuracy by up to 49.26\%. These results confirm the effectiveness of our search engine information enhancement approach.

\begin{figure}[t]  
    \centering  
    \subfloat[]{  
        \includegraphics[width=0.46\linewidth]{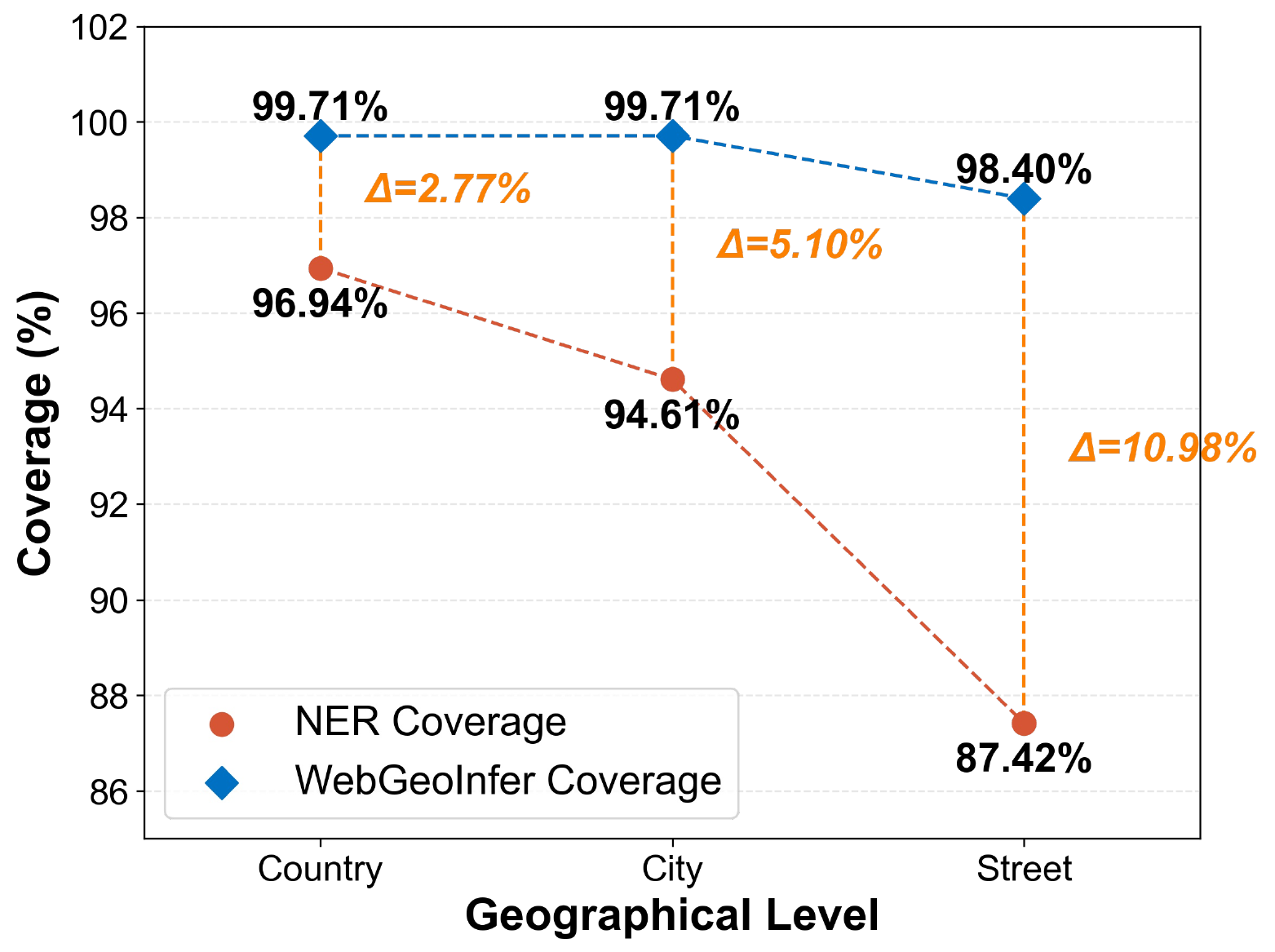}%  
        \label{fig:ner_coverage_country}  
    }  
    \hfil  
    \subfloat[]{  
        \includegraphics[width=0.46\linewidth]{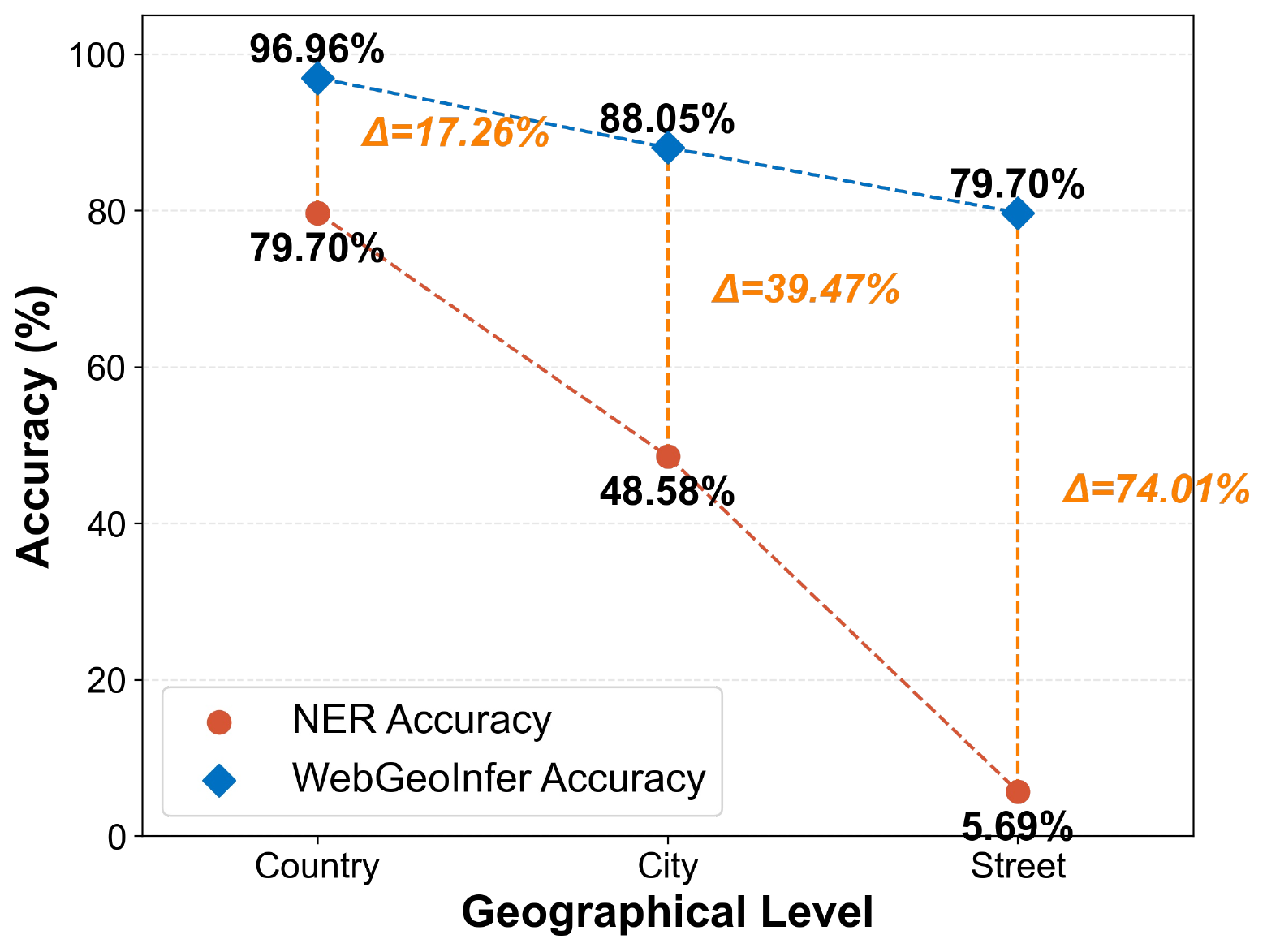}%  
        \label{fig:ner_coverage_city}  
    }  
    \caption{Performance Comparison of NER and Multi-Model Collaborative Inference for Geographic Information Extraction. (a) Coverage at Different Geographic Hierarchy Levels, (b) Accuracy at Different Geographic Hierarchy Levels.}  
    \label{fig:ablation_study_NER}  
\end{figure}

\subsubsection{\textbf{Ablation Study (Comparative Analysis of NER and multi-model collaborative inference)}} To verify the advantages of our multi-model collaborative inference approach compared to NER methods, we compared both approaches' coverage rate and accuracy under identical conditions, using search engine query results to enhance the original differential text. As shown in Figure~\ref{fig:ablation_study_NER}, our method achieves superior performance in coverage rate and accuracy across different levels, particularly in street-level accuracy, where it significantly outperforms NER methods.

\section{Results Analysis}
\label{sec:Results Analysis}

Analysis of remotely managed SCADA device pages from the Censys dataset allows us to infer the geolocation of 5435 devices (2305 via keyword matching + 3130 via LLMs).  We also analyze their geographic distribution to provide a global perspective on their locations.

\subsection{Device Brand Distribution}

In this study, we analyze the brand information of the 5435 devices whose geographic locations were inferred. Due to limited manufacturer information on some web pages, we identify 33 distinct device brands/models. The developers of these devices and their application scenarios span various industrial and commercial fields, with their web interfaces typically containing geographical location data. Table~\ref{tab:scada_devices} displays the various device models and their corresponding industry classifications. The diversity of device types, from energy management to video surveillance systems, determines the varied geographical distribution. For instance, Four-faith's grid automation devices total 967 units, while Obvius's data collection service devices account for 320 units, indicating significant market shares in their respective industries. This confirms the accuracy of our selection of "SCADA" tagged data from the Censys dataset.

\begin{figure*}[t]  
    \centering  
    \subfloat[]{  
        \includegraphics[width=0.47\linewidth]{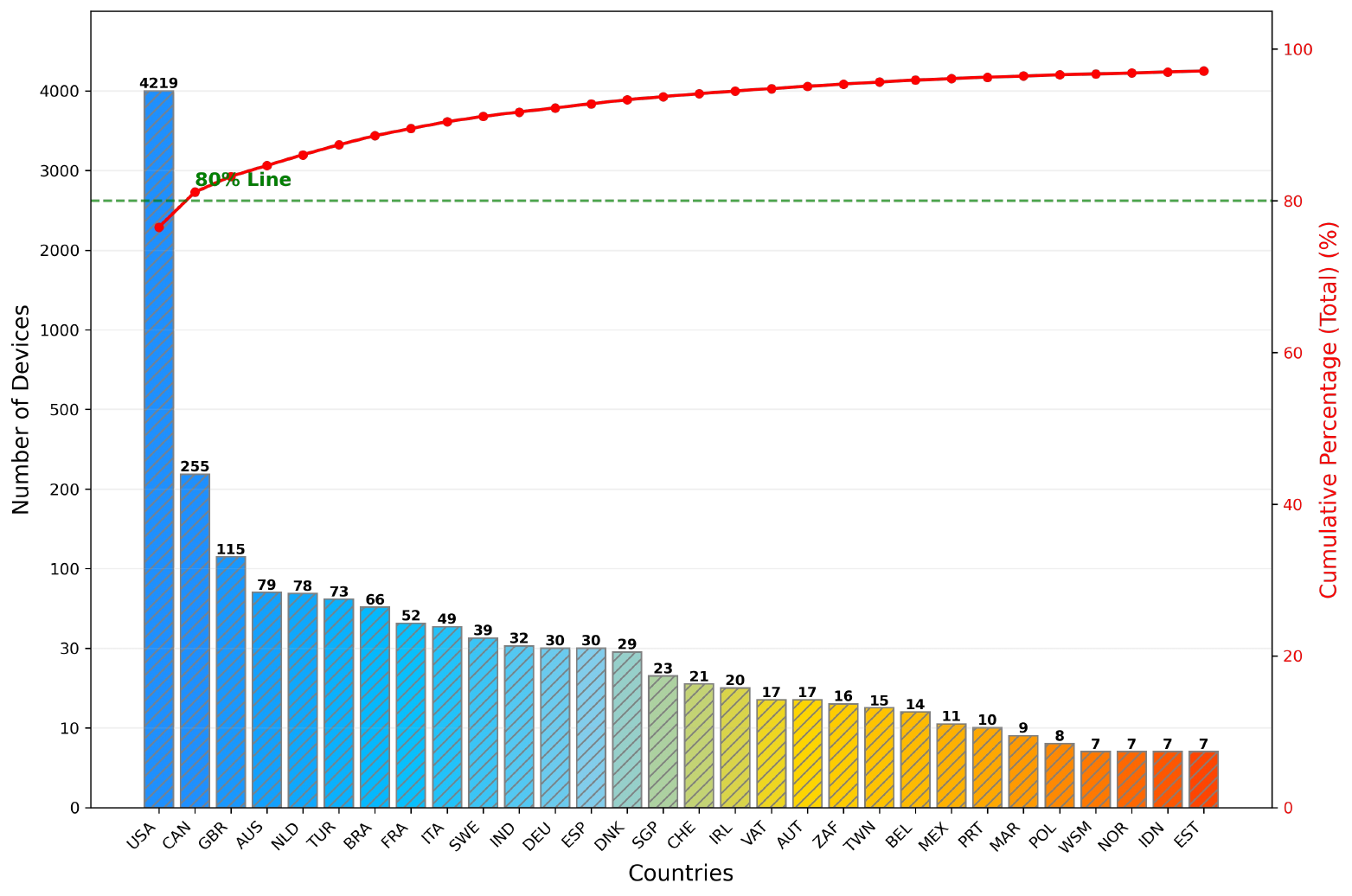}%  
        \label{fig:country_distribution}  
    }  
        \hfil  
    \subfloat[]{  
        \includegraphics[width=0.47\linewidth]{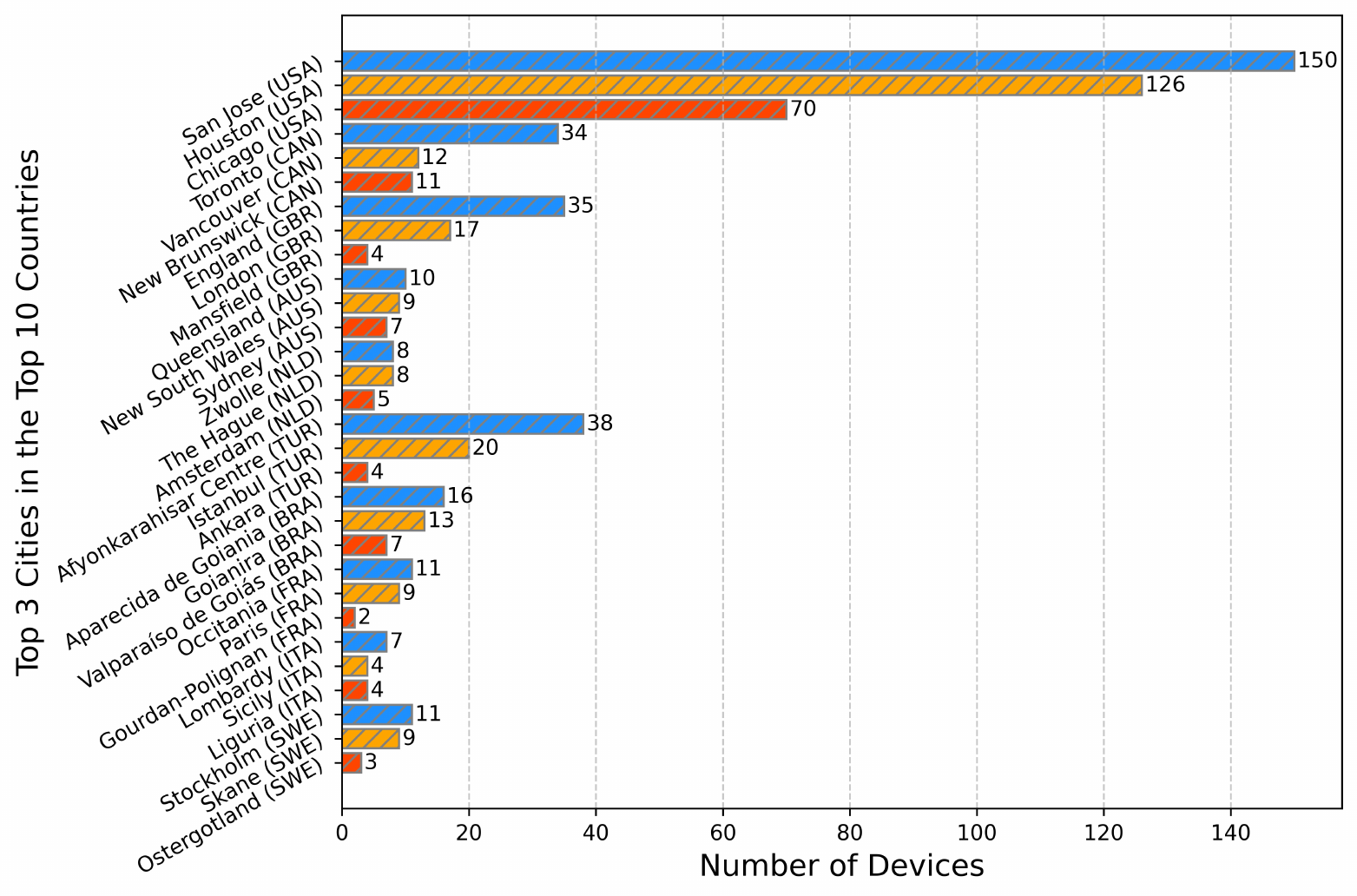}%  
        \label{fig:city_country_distribution}  
    }  
    \caption{Geographic Distribution Analysis of Remote Device Management Inference Results. (a) Country-level Distribution Information, (b) Top 3 Cities with the Most Devices in Top 10 Countries}  
    \label{fig:geo_distribution}  
\end{figure*}

Beyond geographic distribution analysis, our location inference reveals significant security implications. Notably, approximately 38.6\% (2097 out of 5435) of devices utilize the Niagara Web Launcher component. This widespread adoption, combined with geographic clustering (as shown in Fig.~\ref{fig:geo_distribution}), highlights potential supply chain vulnerabilities in the remote management ecosystem. Previous research has identified vulnerabilities in various Niagara framework versions that could affect devices using this component. While specific risk levels depend on implementation details such as version differences and network segmentation, the widespread use of the component increases the theoretical attack surface. Organizations should implement security measures including regular updates, network isolation, and continuous monitoring to mitigate associated risks. This finding demonstrates how geographic inference can contribute to security risk assessment by identifying regional concentrations of potentially vulnerable infrastructure components.

\begin{table}[t]  
    \centering  
    \caption{Distribution of Device Brands and Industries Exposing Geographic Location Information}  
    \label{tab:scada_devices}  
    \resizebox{\columnwidth}{!}{  % 调整表格适应单栏宽度  
    \begin{tabular}{lll>{\raggedleft\arraybackslash}r}  
        \toprule  
        \textbf{Device Model/Brand/Component} & \textbf{Industry} & \textbf{Count} \\
        \midrule  
        Niagara Web Launcher & Web Component & 2097 \\
        Four-faith & Power Grid Automation & 967 \\
        % Unknown & — & 650 \\
        Carrier & Home Air Conditioning & 520 \\
        Obvius & Data Acquisition Services & 320 \\
        XFLOW (Schneider) & Industrial Remote Management & 187 \\
        ASPECT (ABB) & Home \& Building Automation & 128 \\
        SIMATIC HMI & Human-Machine Interface & 102 \\
        Dorian Systems & Building Thermal Control System & 79 \\
        LOYTEC & Controllers & 76 \\
        Kabona & Energy Automation & 54 \\
        Ignition & Industrial Control Software & 43 \\
        Mofi network & Network Management & 43 \\
        Blue Iris & Video Surveillance & 29 \\
        MOBOTIX & Video Surveillance System & 24 \\
        CKSquare & Commercial Manufacturing & 19 \\
        InteLift & Oil Well Management System & 17 \\
        Perle & Network Management & 9 \\
        Telkonet & Energy Management System & 8 \\
        Perax & Remote Management System & 6 \\
        Ipecs & Hybrid Communication Platform & 6 \\
        Digi & Network Management & 6 \\
        NetBiter & Remote Monitoring \& Industrial Management & 5 \\
        Delta Controls & Building Automation Control & 5 \\
        TNT Technologies & Infrastructure & 5 \\
        Rain Bird & Irrigation System Controller & 4 \\
        Avaya & Network Systems & 4 \\
        Dividia & Security Assurance & 4 \\
        Coster Group & Building Energy Management & 3 \\
        RC-WebView & Building Management & 3 \\
        Trihedral & SCADA Software & 3 \\
        Carlo Gavazzi & Photovoltaic Applications & 3 \\
        AMAECO & Energy Management & 3 \\
        PowerLogic (Schneider) & Power Management & 3 \\
        \midrule  
        \textbf{Total} & & \textbf{4785} \\
        \bottomrule  
    \end{tabular}  
    }  
\end{table}

\subsection{Geographic distribution analysis}

As shown in Fig.~\ref{fig:geo_distribution}, we analyze the global geographic distribution of remote management devices, presenting both the country-level overall distribution and the specific distribution across the top 3 cities in the leading 10 countries.

Fig.~\ref{fig:country_distribution} reveals significant distribution disparities at the country level, with the United States notably leading at 4219 devices. Particularly noteworthy is that just the top two countries (United States and Canada) account for 80\% of all devices, forming a classic long-tail distribution pattern. This geographic concentration reflects these nations' first-mover advantages in industrial automation and infrastructure digitalization processes, while also highlighting geopolitical risk concentration points in global SCADA system distribution. From a security perspective, this highly concentrated geographic distribution may constitute significant strategic vulnerabilities under security threat scenarios.

Fig.~\ref{fig:city_country_distribution} further details the city-level distribution within leading countries, analyzing the top 3 cities in each of the top 10 countries. The research finds that devices exhibit clear regional clustering characteristics within each country, primarily congregating in industrial centers, manufacturing hubs, and technology innovation zones. This distribution pattern not only reflects the geographical agglomeration effect of industrial facilities but also confirms the significant correlation between digital infrastructure deployment and urban development levels.

\subsection{AS Distribution Analysis}

\begin{figure}[t]
    \centering
    \includegraphics[width=0.95\linewidth]{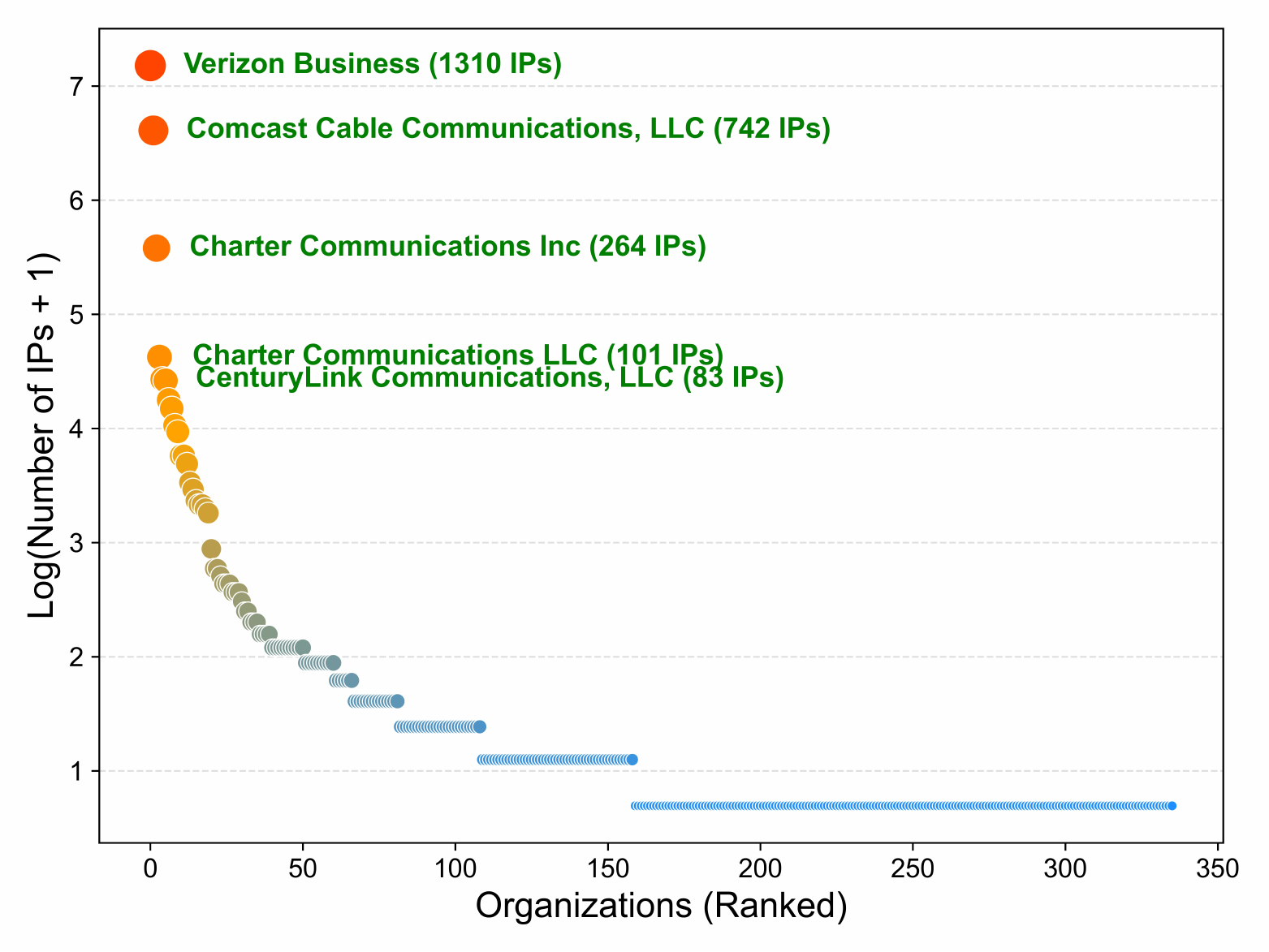}
    \caption{Distribution of Exposed Remote Management Devices Within Top 20 Organizations}
    \label{fig:AS_distribution}
\end{figure}

Beyond geographical distribution, we also analyze the distribution of remote management devices in network topology space. Through IP-AS mapping, we identify network attribution for 3,954 devices, distributed across 336 Autonomous System (AS) organizations.

Our analysis reveals a pronounced long-tail distribution pattern: as shown in Fig.~\ref{fig:AS_distribution}, over 60\% (2500 of 3954) of devices are concentrated in the top 5 ISP networks, while the remaining 40\% are dispersed across nearly 300 ASes of varying sizes. Notably, 53\% of AS organizations host only 1-5 devices each.

This distinctive network distribution pattern reveals multi-layered security challenges:

\textbf{Centralization Risk:} The high concentration of devices within major ISP networks creates ideal scanning environments for attackers. Verizon Business's AS6167, containing over 1,000 exposed devices, exemplifies how attackers can efficiently identify vulnerable targets through targeted scanning, facilitating systematic penetration and scaled attacks.

\textbf{Edge Vulnerability:} Small specialized ISPs serving specific industries (industrial control, healthcare) typically host fewer but high-value devices while possessing insufficient security resources. This creates "invisible weak points" in critical infrastructure. The resulting imbalanced distribution generates a security governance dilemma: while major attack surfaces are concentrated, high-value targets remain scattered across regulatory blind spots.

\begin{figure}[t]
    \centering
    \includegraphics[width=0.95\linewidth]{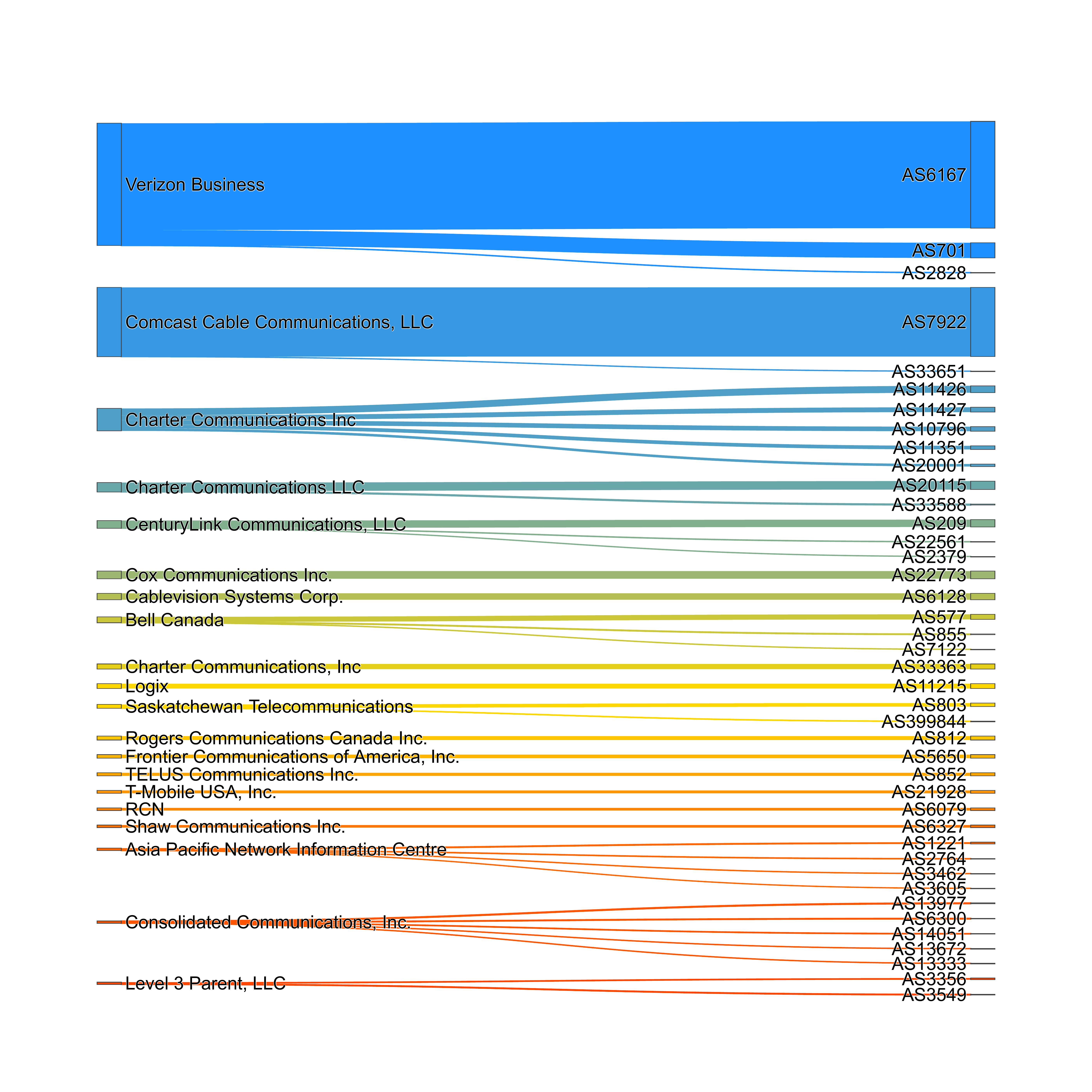}
    \caption{Sankey Diagram of AS to Organization Relationships. The diagram illustrates the mapping between organizations (left) and their corresponding Autonomous System domains (right). The thickness of each connecting line represents the relative number of devices present within each AS domain, providing a visual indication of device distribution across different network infrastructures.}
    \label{fig:Org_as_distribution}
\end{figure}

\textbf{Broken Responsibility Chain:} Device exposure levels within networks directly reflect AS owners' security policy effectiveness. As Fig.~\ref{fig:Org_as_distribution} demonstrates, several organizations have numerous exposed devices across multiple owned ASes, indicating systematic failures to implement effective boundary protection measures.

This distribution pattern simultaneously reflects the internet's hierarchical infrastructure and exposes systemic deficiencies in network security practices. Large ISPs implement "default-open" policies to attract enterprise customers, while smaller ASes lack comprehensive security auditing capabilities. Together, these practices create an uncontrolled exposure surface of management devices on the public internet—highlighting a fundamental collective action problem in current distributed network governance models.

\section{Discussion}
\label{Discussion}

This section addresses the limitations of our approach, explores potential applications in device identification, and examines privacy and ethical considerations.

\subsection{Methodological Limitations}

While our geographic location inference approach has demonstrated effectiveness, several limitations warrant acknowledgment and present opportunities for future research.

Our method targets remote management device web pages containing geographic clues, excluding those without such information, as not all pages reveal location details. When present, our differential text extraction method effectively identifies valuable location indicators. Geographic clue extraction by LLMs depends significantly on search engine information augmentation—a dependency shared by all retrieval-augmented generation systems. To address this limitation, our multi-model collaboration strategy combines inferences from multiple LLMs, enhancing overall system robustness. Geographic information processing faces additional challenges from diverse expression patterns across languages and regions. Variations in address formats, including different writing systems and complex hierarchical structures, present persistent processing difficulties. However, as LLMs' multilingual capabilities continue advancing, we anticipate expanded framework applicability across these diverse contexts.

Additionally, deploying online LLMs involves significant cost considerations. Although open-source offline models are available as alternatives, the costs of large-scale, high-frequency online model API calls pursued for ultimate accuracy cannot be ignored. The optimal solution is to fine-tune a specialized model for geographic information processing in this domain, enabling cost-effective offline deployment.

\subsection{Applications in Fine-Grained Device Identification}

During experimentation, we discovered this framework's potential for device identification applications. The page clustering process in differential text extraction can effectively group web pages from identical manufacturers, while extracted differential texts contain both geographic location and device model information. Once device models are identified, queries through search engines followed by LLM analysis can enable fine-grained recognition of remote management devices, offering novel approaches for network asset identification and security assessment.

\subsection{Privacy and Ethical Considerations}

Although this research involves inferring the geographic locations of remote management devices, our data are entirely sourced from publicly accessible resources, which significantly reduces ethical risks. These publicly available data do not constitute protected information, minimizing the risk of unauthorized surveillance or targeted attacks. Simultaneously, our primary research objective is to demonstrate the feasibility of geographic inference through device web pages to enhance regulatory measures. To prevent geographic information exposure on public internet platforms, we recommend the following mitigation strategies:

\begin{itemize}
    \item \textbf{ISP Perspective}: Strengthen compliance and oversight frameworks to ensure user privacy protection, including adherence to data protection regulations and implementation of industry best practices.
    \item \textbf{Device Manufacturer Perspective}: Develop systems that guide users to minimize sensitive information exposure when configuring critical fields, and implement privacy-protecting default configurations.
    \item \textbf{Device Administrator Perspective}: Establish robust access control policies that restrict geographic location information access; enhance security awareness to minimize sensitive information exposure during device configuration.
\end{itemize}

\section{Conclusion}
\label{Conclusion}

We propose a remote management device geographic location inference framework called WebGeoInfer, featuring two innovations: a differential text extraction method that identifies content variations between device web pages without relying on web page structure, and a multi-model collaborative analysis technique combining search engines with language models to extract location information from implicit text.

Our framework successfully inferred device locations globally with high accuracy across country, city, and street levels. Through analyzing geographic and AS distribution patterns, we revealed key deployment characteristics and security vulnerabilities, contributing valuable insights to network security research and assessment.

\bibliographystyle{IEEEtran}
\bibliography{reference}

\begin{thebibliography}{10}
\providecommand{\url}[1]{#1}
\csname url@samestyle\endcsname
\providecommand{\newblock}{\relax}
\providecommand{\bibinfo}[2]{#2}
\providecommand{\BIBentrySTDinterwordspacing}{\spaceskip=0pt\relax}
\providecommand{\BIBentryALTinterwordstretchfactor}{4}
\providecommand{\BIBentryALTinterwordspacing}{\spaceskip=\fontdimen2\font plus
\BIBentryALTinterwordstretchfactor\fontdimen3\font minus \fontdimen4\font\relax}
\providecommand{\BIBforeignlanguage}[2]{{%
\expandafter\ifx\csname l@#1\endcsname\relax
\typeout{** WARNING: IEEEtran.bst: No hyphenation pattern has been}%
\typeout{** loaded for the language `#1'. Using the pattern for}%
\typeout{** the default language instead.}%
\else
\language=\csname l@#1\endcsname
\fi
#2}}
\providecommand{\BIBdecl}{\relax}
\BIBdecl

\bibitem{sasaki2022exposed}
T.~Sasaki, A.~Fujita, C.~H. Gan{\'a}n, M.~van Eeten, K.~Yoshioka, and T.~Matsumoto, ``Exposed infrastructures: Discovery, attacks and remediation of insecure ics remote management devices,'' in \emph{2022 IEEE Symposium on Security and Privacy (SP)}.\hskip 1em plus 0.5em minus 0.4em\relax IEEE, 2022, pp. 2379--2396.

\bibitem{liu2022discover}
J.~Liu, J.~Wang, P.~Liu, H.~Zhu, and L.~Sun, ``Discover the ics landmarks based on multi-stage clue mining,'' in \emph{International Conference on Wireless Algorithms, Systems, and Applications}.\hskip 1em plus 0.5em minus 0.4em\relax Springer, 2022, pp. 139--151.

\bibitem{zilberman2024survey}
A.~Zilberman, A.~Offer, B.~Pincu, Y.~Glickshtein, R.~Kant, O.~Brodt, A.~Otung, R.~Puzis, A.~Shabtai, and Y.~Elovici, ``A survey on geolocation on the internet,'' \emph{IEEE Communications Surveys \& Tutorials}, 2024.

\bibitem{ma2023graphnei}
Z.~Ma, S.~Zhang, N.~Li, T.~Li, X.~Hu, H.~Feng, Q.~Zhou, F.~Liu, X.~Quan, H.~Wang \emph{et~al.}, ``Graphnei: A gnn-based network entity identification method for ip geolocation,'' \emph{Computer Networks}, vol. 235, p. 109946, 2023.

\bibitem{liu2025gdd}
C.~Liu, R.~Cheng, F.~Yuan, S.~Ding, Y.~Liu, and X.~Luo, ``Gdd-geo: Ipv6 geolocation by graph dual decomposition,'' \emph{Computer Communications}, vol. 231, p. 108019, 2025.

\bibitem{cho2024selection}
S.~Cho, Z.~Weinberg, A.~Bhattacharya, S.~Dai, and R.~Rauf, ``Selection of landmarks for efficient active geolocation,'' in \emph{2024 8th Network Traffic Measurement and Analysis Conference (TMA)}.\hskip 1em plus 0.5em minus 0.4em\relax IEEE, 2024, pp. 1--9.

\bibitem{zu2022ip}
S.~Zu, X.~Luo, and F.~Zhang, ``Ip-geolocater: a more reliable ip geolocation algorithm based on router error training,'' \emph{Frontiers of Computer Science}, vol.~16, no.~1, p. 161504, 2022.

\bibitem{lin2024probegeo}
J.~Lin, C.~Li, W.~Gong, G.~Song, L.~Fan, Z.~Wang, and J.~Yang, ``Probegeo: A comprehensive landmark mining framework based on web content,'' \emph{IEEE/ACM Transactions on Networking}, 2024.

\bibitem{yan2022internet}
Z.~Yan, Z.~Li, H.~Li, S.~Yang, H.~Zhu, and L.~Sun, ``Internet-scale fingerprinting the reusing and rebranding iot devices in the cyberspace,'' \emph{IEEE Transactions on Dependable and Secure Computing}, vol.~20, no.~5, pp. 3890--3909, 2022.

\bibitem{cheng2021identify}
H.~Cheng, W.~Dong, Y.~Zheng, and B.~Lv, ``Identify iot devices through web interface characteristics,'' in \emph{2021 IEEE 6th International Conference on Computer and Communication Systems (ICCCS)}.\hskip 1em plus 0.5em minus 0.4em\relax IEEE, 2021, pp. 405--410.

\bibitem{dan2022ip}
O.~Dan, V.~Parikh, and B.~D. Davison, ``Ip geolocation through geographic clicks,'' \emph{ACM Transactions on Spatial Algorithms and Systems (TSAS)}, vol.~8, no.~1, pp. 1--22, 2022.

\bibitem{gueye2004constraint}
B.~Gueye, A.~Ziviani, M.~Crovella, and S.~Fdida, ``Constraint-based geolocation of internet hosts,'' in \emph{Proceedings of the 4th ACM SIGCOMM conference on Internet measurement}, 2004, pp. 288--293.

\bibitem{laki2011spotter}
S.~Laki, P.~M{\'a}tray, P.~H{\'a}ga, T.~Seb{\H{o}}k, I.~Csabai, and G.~Vattay, ``Spotter: A model based active geolocation service,'' in \emph{2011 Proceedings IEEE INFOCOM}.\hskip 1em plus 0.5em minus 0.4em\relax IEEE, 2011, pp. 3173--3181.

\bibitem{wang2011towards}
Y.~Wang, D.~Burgener, M.~Flores, A.~Kuzmanovic, and C.~Huang, ``Towards $\{$Street-Level$\}$$\{$Client-Independent$\}$$\{$IP$\}$ geolocation,'' in \emph{8th USENIX Symposium on Networked Systems Design and Implementation (NSDI 11)}, 2011.

\bibitem{xiang2019no}
C.~Xiang, X.~Wang, Q.~Chen, M.~Xue, Z.~Gao, H.~Zhu, C.~Chen, and Q.~Fan, ``No-jump-into-latency in china's internet! toward last-mile hop count based ip geo-localization,'' in \emph{Proceedings of the International Symposium on Quality of Service}, 2019, pp. 1--10.

\bibitem{katz2006towards}
E.~Katz-Bassett, J.~P. John, A.~Krishnamurthy, D.~Wetherall, T.~Anderson, and Y.~Chawathe, ``Towards ip geolocation using delay and topology measurements,'' in \emph{Proceedings of the 6th ACM SIGCOMM conference on Internet measurement}, 2006, pp. 71--84.

\bibitem{ding2023gnn}
S.~Ding, X.~Luo, J.~Wang, and X.~Fu, ``Gnn-geo: a graph neural network-based fine-grained ip geolocation framework,'' \emph{IEEE Transactions on Network Science and Engineering}, vol.~10, no.~6, pp. 3543--3560, 2023.

\bibitem{wang2022connecting}
Z.~Wang, F.~Zhou, W.~Zeng, G.~Trajcevski, C.~Xiao, Y.~Wang, and K.~Chen, ``Connecting the hosts: Street-level ip geolocation with graph neural networks,'' in \emph{Proceedings of the 28th ACM SIGKDD conference on knowledge discovery and data mining}, 2022, pp. 4121--4131.

\bibitem{wang2025neighborgeo}
X.~Wang, D.~Zhao, X.~Liu, Z.~Zhang, and T.~Zhao, ``Neighborgeo: Ip geolocation based on neighbors,'' \emph{Computer Networks}, vol. 257, p. 110896, 2025.

\bibitem{mansoori2020they}
M.~Mansoori and I.~Welch, ``How do they find us? a study of geolocation tracking techniques of malicious web sites,'' \emph{Computers \& Security}, vol.~97, p. 101948, 2020.

\bibitem{berragan2023transformer}
C.~Berragan, A.~Singleton, A.~Calafiore, and J.~Morley, ``Transformer based named entity recognition for place name extraction from unstructured text,'' \emph{International Journal of Geographical Information Science}, vol.~37, no.~4, pp. 747--766, 2023.

\bibitem{cruz2013semantic}
I.~F. Cruz, V.~R. Ganesh, and S.~I. Mirrezaei, ``Semantic extraction of geographic data from web tables for big data integration,'' in \emph{Proceedings of the 7th Workshop on Geographic Information Retrieval}, 2013, pp. 19--26.

\bibitem{endo2010whois}
P.~T. Endo and D.~F.~H. Sadok, ``Whois based geolocation: A strategy to geolocate internet hosts,'' in \emph{2010 24th IEEE International Conference on Advanced Information Networking and Applications}.\hskip 1em plus 0.5em minus 0.4em\relax IEEE, 2010, pp. 408--413.

\bibitem{corneo2024whois}
L.~Corneo and M.~Di~Francesco, ``From whois to rdap: Are ip lookup services getting any better?'' in \emph{NOMS 2024-2024 IEEE Network Operations and Management Symposium}.\hskip 1em plus 0.5em minus 0.4em\relax IEEE, 2024, pp. 1--10.

\bibitem{gasser2016scanning}
O.~Gasser, Q.~Scheitle, S.~Gebhard, and G.~Carle, ``Scanning the ipv6 internet: towards a comprehensive hitlist,'' \emph{arXiv preprint arXiv:1607.05179}, 2016.

\bibitem{komosny2017location}
D.~Komosny, M.~Voznak, and S.~U. Rehman, ``Location accuracy of commercial ip address geolocation databases,'' \emph{Information technology and control}, vol.~46, no.~3, pp. 333--344, 2017.

\bibitem{rye2023ipvseeyou}
E.~C. Rye and R.~Beverly, ``Ipvseeyou: Exploiting leaked identifiers in ipv6 for street-level geolocation,'' in \emph{2023 IEEE Symposium on Security and Privacy (SP)}.\hskip 1em plus 0.5em minus 0.4em\relax IEEE, 2023, pp. 3129--3145.

\bibitem{padmanabhan2001investigation}
V.~N. Padmanabhan and L.~Subramanian, ``An investigation of geographic mapping techniques for internet hosts,'' in \emph{Proceedings of the 2001 conference on Applications, technologies, architectures, and protocols for computer communications}, 2001, pp. 173--185.

\bibitem{huffaker2014drop}
B.~Huffaker, M.~Fomenkov, and K.~Claffy, ``Drop: Dns-based router positioning,'' \emph{ACM SIGCOMM Computer Communication Review}, vol.~44, no.~3, pp. 5--13, 2014.

\bibitem{dan2021ip}
O.~Dan, V.~Parikh, and B.~D. Davison, ``Ip geolocation through reverse dns,'' \emph{ACM Transactions on Internet Technology (TOIT)}, vol.~22, no.~1, pp. 1--29, 2021.

\bibitem{mun2017building}
H.~Mun and Y.~Lee, ``Building ip geolocation database from online used market articles,'' in \emph{2017 19th Asia-Pacific Network Operations and Management Symposium (APNOMS)}.\hskip 1em plus 0.5em minus 0.4em\relax IEEE, 2017, pp. 37--41.

\bibitem{du2019geoblr}
F.~Du, X.~Bao, Y.~Zhang, and Y.~Wang, ``Geoblr: Dynamic ip geolocation method based on bayesian linear,'' in \emph{Collaborative Computing: Networking, Applications and Worksharing: 14th EAI International Conference, CollaborateCom 2018, Shanghai, China, December 1-3, 2018, Proceedings}, vol. 268.\hskip 1em plus 0.5em minus 0.4em\relax Springer, 2019, p. 310.

\bibitem{wang2019one}
Y.~Wang, X.~Wang, H.~Zhu, H.~Zhao, H.~Li, and L.~Sun, ``One-geo: client-independent ip geolocation based on owner name extraction,'' in \emph{Wireless Algorithms, Systems, and Applications: 14th International Conference, WASA 2019, Honolulu, HI, USA, June 24--26, 2019, Proceedings 14}.\hskip 1em plus 0.5em minus 0.4em\relax Springer, 2019, pp. 346--357.

\bibitem{chang2024survey}
Y.~Chang, X.~Wang, J.~Wang, Y.~Wu, L.~Yang, K.~Zhu, H.~Chen, X.~Yi, C.~Wang, Y.~Wang \emph{et~al.}, ``A survey on evaluation of large language models,'' \emph{ACM transactions on intelligent systems and technology}, vol.~15, no.~3, pp. 1--45, 2024.

\bibitem{harrod2024text}
K.~Harrod, P.~Bhandari, and A.~Anastasopoulos, ``From text to maps: Llm-driven extraction and geotagging of epidemiological data,'' in \emph{Proceedings of the Third Workshop on NLP for Positive Impact}, 2024, pp. 258--270.

\bibitem{chen2025ipdb}
H.~Chen, G.~Song, Z.~Wang, J.~Yang, S.~Wu, J.~Lin, L.~He, and C.~Li, ``Ipdb: A high-precision ip level industry categorization of web services,'' in \emph{THE WEB CONFERENCE 2025}.

\bibitem{du2020ripe}
B.~Du, M.~Candela, B.~Huffaker, A.~C. Snoeren, and K.~Claffy, ``Ripe ipmap active geolocation: Mechanism and performance evaluation,'' \emph{ACM SIGCOMM Computer Communication Review}, vol.~50, no.~2, pp. 3--10, 2020.

\bibitem{gharaibeh2017look}
M.~Gharaibeh, A.~Shah, B.~Huffaker, H.~Zhang, R.~Ensafi, and C.~Papadopoulos, ``A look at router geolocation in public and commercial databases,'' in \emph{Proceedings of the 2017 Internet Measurement Conference}, 2017, pp. 463--469.

\bibitem{electronics13010015}
\BIBentryALTinterwordspacing
Y.~Xie, Z.~Zhang, Y.~Liu, E.~Chen, and N.~Li, ``Evaluation method of ip geolocation database based on city delay characteristics,'' \emph{Electronics}, vol.~13, no.~1, 2024. [Online]. Available: \url{https://www.mdpi.com/2079-9292/13/1/15}
\BIBentrySTDinterwordspacing

\bibitem{censys15}
Z.~Durumeric, D.~Adrian, A.~Mirian, M.~Bailey, and J.~A. Halderman, ``A search engine backed by {I}nternet-wide scanning,'' in \emph{22nd {ACM} Conference on Computer and Communications Security}, Oct. 2015.

\end{thebibliography}

\end{document}